\documentclass[sigconf]{acmart}
\usepackage{listings}
\usepackage{balance}
\usepackage[justification=centerlast, skip=12pt]{caption}
\usepackage{appendix}
\usepackage{color}
\usepackage{soul}
\usepackage{balance}
\usepackage{multirow}

\newcommand{\minisec}[1]{\noindent\textbf{#1.}}

\newcommand{\minisecnodot}[1]{\noindent\textbf{#1}}
\newcommand{\reactdb}{\textsc{ReactDB}}

\newcommand{\techreport}[1]{}

\begin{document}
\copyrightyear{2018}
\acmYear{2018}
\setcopyright{acmcopyright}
\acmConference[SIGMOD'18]{2018 International Conference on
Management of Data}{June 10--15, 2018}{Houston, TX, USA}
\acmBooktitle{SIGMOD'18: 2018 International Conference on
Management of Data, June 10--15, 2018, Houston, TX, USA}
\acmPrice{15.00}
\acmDOI{10.1145/3183713.3183752}
\acmISBN{978-1-4503-4703-7/18/06}
\settopmatter{printacmref=false}
\fancyhead{}

\title{Reactors: A Case for Predictable, Virtualized Actor
  Database Systems}

\subtitle{\textit{Minor Revision}}
\subtitlenote{The changes from the original version~\cite{ShahS18:Reactors} consist of minor reformatting and an expanded appendix with additional experiments, figures and code examples.}

\author{Vivek Shah}
\affiliation{
\institution{University of Copenhagen, Denmark}
}
\email{bonii@di.ku.dk}
\author{Marcos Antonio Vaz Salles}
\affiliation{
\institution{University of Copenhagen, Denmark}
}
\email{vmarcos@di.ku.dk}
\begin{abstract}
The requirements for OLTP database systems are becoming ever more
demanding. Domains such as finance and computer games increasingly
mandate that developers be able to encode complex application
logic and control transaction latencies in in-memory databases. At
the same time, infrastructure engineers in these domains need to
experiment with and deploy OLTP database architectures that ensure
application scalability and maximize resource utilization in
modern machines. In this paper, we propose a relational actor programming model for in-memory databases as a novel, holistic approach towards fulfilling these challenging requirements. Conceptually, relational actors, or \emph{reactors} for short, are application-defined, isolated logical actors that encapsulate relations and process function calls asynchronously. Reactors ease reasoning about correctness by guaranteeing serializability of application-level function calls. In contrast to classic transactional models, however, reactors allow developers to take advantage of intra-transaction parallelism and state encapsulation in their applications to reduce latency and improve locality. Moreover, reactors enable a new degree of flexibility in database deployment. We present \reactdb, a system design exposing reactors that allows for flexible virtualization of database architecture between the extremes of shared-nothing and shared-everything without changes to application code. Our experiments illustrate latency control, low overhead, and asynchronicity trade-offs with \reactdb\ in OLTP benchmarks.
\end{abstract}
\maketitle

\section{Introduction}
\label{sec:intro}

Three trends are transforming the landscape of OLTP systems. First, a host of latency-sensitive OLTP applications has emerged in areas as diverse as computer games, high-performance trading, and the web~\cite{Brook15:Finance,Stonebraker12:NewSQL,WhiteKGGD07:Games}. This trend brings about challenging performance requirements, including mechanisms to allow developers to reason about transaction latencies and scalability of their applications with large data and request volumes~\cite{ShoupPl06:eBay,Pattishall07:Flickr}. Second, database systems are moving towards solid state, in particular in-memory storage~\cite{DEB:MainMemory}, and hardware systems are integrating increasingly more cores in a single machine. This trend brings about new requirements for database architecture, such as processing efficiency in multi-core machines and careful design of concurrency control strategies~\cite{Tu:2013:STM:2517349.2522713,YuBPDS14:ThousandCores}.  Third, there is a need to operate databases out of virtualized infrastructures with high resource efficiency~\cite{Bernstein:2011:SQLAzure,KossmannKL10:EvaluationCloudDB,NarasayyaDSCC13:SQLVM}. This trend leads to the requirement that virtualization abstractions for databases impose low overhead and allow for flexible deployments without causing changes to application~programs. 

Recent research in OLTP databases has shown that addressing all of these requirements is a hard problem. 
On the one hand, shared-nothing database designs, such as those of H-Store~\cite{Stonebraker:2007:EAE:1325851.1325981} or HyPer~\cite{Kemper:2011:HHO:2004686.2005619}, fail to provide appropriately for multi-core efficiency in the presence of cross-partition transactions~\cite{Diaconu:2013:HSS:2463676.2463710,Tu:2013:STM:2517349.2522713}. This is due to the impact of overheads in mapping partitions to cores and of synchronous communication in distributed transactions across partitions. Consequently, transaction  throughput and latencies in these systems are very sensitive to how data is partitioned. 
On the other hand, shared-everything databases have a hard time achieving multi-core scalability. To do so, these systems either internally partition their data structures, e.g., DORA~\cite{Pandis:2010:DTE:1920841.1920959}, or benefit from affinity of memory accesses to cores in transactions, 
e.g., Silo~\cite{Tu:2013:STM:2517349.2522713}. Thus, deployment decisions can affect efficiency and scalability in these systems and are difficult to get right across application~classes. 

As a consequence, both developers and infrastructure engineers in demanding OLTP domains have a hard time controlling the performance of their transactional databases. Despite advances in profiling tools to identify causes of latency variance in database systems~\cite{HuangMSW17:TProfiler}, today developers lack clear abstractions to reason at a high level about the interplay of complex, potentially parallelizable application logic and transaction latencies. In addition, the variety of modern in-memory database engines, including numerous specialized designs ranging internally from shared-nothing to shared-everything~\cite{YaoA0LOWZ16:Lads,RenFA16:Orthrus,Pandis:2011:PPL:2021017.2021019}, challenges the ability of infrastructure engineers to flexibly experiment with and adapt database architecture without affecting application code. 

Actor programming models provide desirable primitives for concurrent and distributed programming~\cite{Agha:1986:Actors, Armstrong10:Erlang, HaydukSF15:STMActor}, which of late have evoked a strong interest in the database community~\cite{BernsteinDKM17:ActorDB}. 
To holistically meet the challenging requirements imposed on OLTP systems, we propose a new actor programming model in relational databases called \emph{Relational Actors} (or \emph{reactors} for short). Reactors are special types of actors that model logical computational entities encapsulating state abstracted as relations. For example, reactors can represent application-level scaling units such as accounts in a banking application or warehouses in a retail management application. 
Within a reactor, developers can take advantage of classic database programming features such as declarative querying over the encapsulated relations. To operate on state logically distributed across reactors, however, developers employ explicit asynchronous function calls. The latter allows developers of latency-sensitive OLTP applications to write their programs so as to minimize cross-reactor accesses or overlap communication with computation. 
 Still, a transaction across multiple reactors provides serializability guarantees as in traditional databases, thus relieving developers from struggling with complex concurrency issues.
 Reactors allow application-level modeling between the extremes of a relational database (single reactor encapsulating all relations) and key-value stores (each reactor encapsulating a key-value pair).

To address the challenges of architectural flexibility and high resource efficiency in multi-core machines, we design an in-memory database system that exposes reactors as a programming model. This system, called \reactdb\ (RElational ACTor DataBase), 
decomposes and virtualizes the notions of sharing of compute and memory in database architecture. First, we introduce a database containerization scheme to enclose shared-memory regions in a machine,  each storing state for one or many reactors. Second, within a container, compute resources abstracted as transaction executors can be deployed to either share or own reactors. The combination of these two notions allows infrastructure engineers to experiment with deployments capturing a range of database architecture patterns by simply changing a configuration file. At the same time, no changes are required to application code using reactors.  

\begin{figure*}[!t]
        \begin{minipage}{1.0\linewidth}
          \centerline{\includegraphics[width=0.85\linewidth]{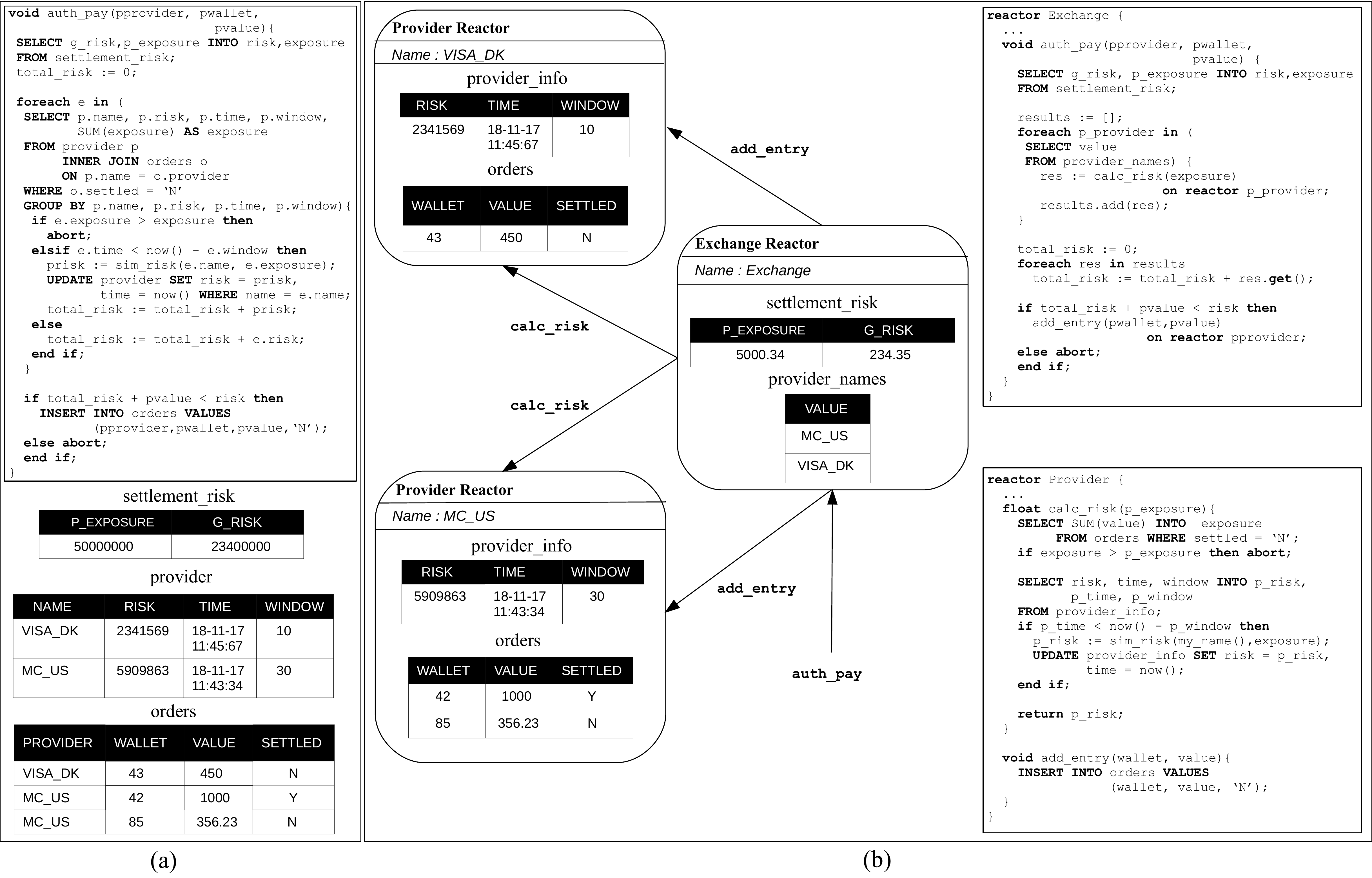}}
        \end{minipage} \hfill
\caption{A simplified currency exchange application in: (a)~the classic transactional model, and (b)~the reactor model.} 
\label{fig:example:app}
\end{figure*}

\minisec{Example: Digital Currency Exchange}
We abstract an application coded using a set of reactors as a \emph{reactor database}. Consider a simplified digital currency exchange application, in which users may buy or sell currency through their credit card providers. Figure~\ref{fig:example:app} contrasts how such an application would be written with a classic transactional database and a reactor database in parts~(a) and~(b), respectively. The exchange wishes to limit its settlement risk from cancelled credit card payments. To do so, it follows two application rules: (1)~stop accepting orders if any single provider's total unsettled exposure goes above the \texttt{p\_exposure} threshold in relation \texttt{settlement\_risk}; (2)~reject orders that cause the total risk-adjusted exposure across all providers to exceed the \texttt{g\_risk} threshold. In part~(a), this logic is encoded in a stored procedure that computes risk-adjusted exposure through an expensive function \texttt{sim\_risk} and caches its result for a time period.    
In part~(b), the same logic is expressed with reactors. The exchange and each of the providers are modeled as relational actors with private state (relations) that can execute certain procedures. The \texttt{Exchange} reactor can execute \texttt{auth\_pay} and encapsulates information about provider names and the settlement limits. \texttt{Provider} reactors store information from risk-adjusted exposure calculations per provider as well as fragments of the relation \texttt{orders} with the payments for each provider, and 
can execute procedures \texttt{calc\_risk} and \texttt{add\_entry}. In \texttt{auth\_pay}, the \texttt{Exchange} reactor invokes asynchronous calls to \texttt{Provider} reactors, making explicit the available intra-transaction parallelism. Since the exchange strives for the lowest latency possible, this program formulation improves transaction response time with respect to part~(a) in a way that is clearly explainable to a developer pursuing application performance optimization. In addition, it becomes explicit in transaction programs that code is conceptually moved close to the data it touches, allowing developers to control for locality. 
At the same time, ACID properties are guaranteed for \texttt{auth\_pay}, despite asynchronous calls to a function, \texttt{calc\_risk}, that includes database updates, user-defined abort conditions, and potentially even nondeterminism in the calculation of \texttt{sim\_risk}~\cite{HaasJAXPJ10:MCDB-R}.  
We further elaborate on this example and our model in Section~\ref{sec:model}. 

\minisecnodot{How are reactors different from database partitioning?}\\
In contrast to database partitioning, which is a data-oriented optimization, reactors represent a \emph{compute-oriented abstraction}. In the example above, reactors can be used to model horizontal partitioning of tuples, vertical partitioning of relations, or any arbitrary grouping of relation fragments. In addition, reactors allow for functional decomposition and modeling of affinity and parallelism in arbitrary application logic. In the example, detecting and fully exploiting the parallelism of \texttt{auth\_pay} in the stored procedure formulation of part~(a) in a classic relational database would require complex control-flow analysis, and may not be possible at~all~\cite{postgresql-parallel-safety}.

\minisec{Contributions} In short, we make the following contributions:

\begin{enumerate}

\item We present a novel logical abstraction for relational databases called \emph{reactors}. This abstraction is grounded on transactional semantics offering serializability and an asynchronous programming model that allows for \emph{encoding complex application logic while considering relative latency of different transactional programs} 
(Section~\ref{sec:model}).

\item We discuss the design of \reactdb, an in-memory database system exposing reactors. \reactdb\ enables \emph{configuration of database architecture at deployment time} in a multi-core machine without changes to application code (Section~\ref{sec:sys:arch}).

\item In experiments with classic OLTP benchmarks, reactors provide latency control at the microsecond scale for varied program formulations. In addition, for given program formulations, database architecture can be configured to control the trade-off between asynchronicity and load (Section~\ref{sec:evaluation}).       
 
\end{enumerate}

\section{Programming Model}
\label{sec:model}
\techreport{In this section, we introduce the reactor programming model. First, the main concepts of reactors are outlined in Section~\ref{sec:programming:model:concepts}, before a detailed explanation of the programming model in Section~\ref{sec:programming:model}. Section~\ref{sec:math:formalism} formalizes conflict-serializability of transactions with reactors. Section~\ref{sec:perf:predict} then describes how reactors allow for reasoning about performance of programs.}

\subsection{Reactor Concepts}
In contrast to classic transactional or actor models, reactors bring together all of the following concepts:
\label{sec:programming:model:concepts}
\begin{enumerate}
\item A reactor is an application-defined logical actor that encapsulates state abstracted using relations.
\item Declarative queries are supported only on a single reactor. Communication across reactors is achieved by asynchronous function calls. A computation (function) across reactors consists of a sequence of intra-reactor statements and/or nested cross-reactor function calls.
\item Computations across reactors have transactional guarantees.
\item Reactors provide an abstract computational cost model for reflecting on relative latency across program formulations.
\end{enumerate}

\subsection{Programming with Reactors}
\label{sec:programming:model}
\subsubsection{Application-Defined Relational Actors}
A \emph{reactor} is an actor specialized for the management of state abstracted by the relational model. The pseudocode in Figure~\ref{fig:txn:int:actor} conceptualizes the capabilities of a reactor. As a regular actor~\cite{Agha:1986:Actors}, a reactor encapsulates a state, which can be accessed by computations invoked on the reactor. However, unlike in a regular actor, in which communication is typically achieved by non-blocking send and blocking receive primitives, the only form of communication with a reactor is through \emph{asynchronous function calls} returning promises~\cite{Liskov:1988:PLS:53990.54016}. Moreover, the code of such functions is similar to that of database stored procedures, which can intermix declarative queries over relations with other program logic and function calls. These asynchronous function calls are abstracted in Figure~\ref{fig:txn:int:actor} by the \texttt{execute} function, which takes as an argument a function to be computed on the reactor's state along with appropriate arguments, and returns a promise representing the result of the computation. In the remainder of this paper, we refer to such a result as a \emph{future}, and use the terms \emph{function} and \emph{procedure} on a reactor interchangeably. 

\begin{figure}
\centering
\begin{lstlisting}
Reactor : Actor {
    RelationalState rs;

    Future execute(compute_fn, args) {
        return new Future(compute_fn(args,rs));
    }
}
\end{lstlisting}
\caption{Conceptual view of reactors as an actor type.}
\label{fig:txn:int:actor}
\vspace{-4ex}
\end{figure}

A \emph{reactor database} is a collection of reactors, each of which must abide by a reactor type. Reactor types, defined by application developers, specify the functions that should be invoked in a reactor and determine the relation schemas encapsulated in the reactor state. To instantiate a reactor database, we need to declare the names and types of the reactors constituting the database and a schema creation function for each reactor type. 
The developer cannot create or destroy reactors; these purely logical entities are accessible by their declared names for the lifetime of the application, bound by the failure model of the reactor database. In contrast to objects in an object-oriented database~\cite{LecluseR89:O2,CareyDV88:Exodus}, reactors are active computational entities, i.e., a reactor is a combination of a logical thread of control and an encapsulated relational state accessible \emph{exclusively} by that logical thread. While objects encapsulate complex types, reactors encapsulate whole relational schemas; declarative querying happens only within, not across reactors, and communication across reactors is explicit through asynchronous function calls.

In the example of Figure~\ref{fig:example:app}(b), the state of a \texttt{Provider} reactor results from both horizontal and vertical fragmentation of the original \texttt{providers} relation from part~(a) as well as horizontal fragmentation of the \texttt{orders} relation (with omission of the \texttt{provider} column). The state of the \texttt{Exchange} reactor retains a relation \texttt{provider\_names}, since the latter is necessary for accessing \texttt{Provider} reactors, and the \texttt{settlement\_risk} relation. This illustrates that different reactors may contain either the same or different schemas, according to their types. 

It is not necessary to know in advance all the providers and their names to model the reactor database. It is sufficient to know: (1)~the types of the reactors expected, namely \texttt{Exchange} and \texttt{Provider}; 
(2)~the schemas and functions of each reactor type; (3)~the name mapping to address provider reactors. As such, adding new providers does not necessitate rewriting the application logic.

\subsubsection{Asynchronous Function Calls}
\label{sec:reactor:async}
To invoke a procedure on a reactor, we must explicitly use the declared name of the reactor where the computation must be executed. The procedure logic can access the relational state on the reactor where it is invoked through declarative queries. If the procedure needs to access the state of another reactor, then it must invoke another procedure on the target reactor. This is necessary because the states of different reactors are disjoint. Since the result of a procedure is represented by a future, the calling code can choose to wait for the result of the future, invoke procedures on other reactors, or execute further application logic. This flexibility allows application developers to expose parallelism within the procedure. 

In Figure~\ref{fig:example:app}(b), the syntax \texttt{procedure\_name(args) \textbf{on reactor} reactor\_name} specifies an asynchronous procedure call routed to the reactor with a given name. In the logic of \texttt{auth\_pay}, the execution of \texttt{calc\_risk} is overlapped on each of the \texttt{Provider} reactors and the futures returned by the calls are stored in the \texttt{results} list. 
The \texttt{Exchange} reactor sums up the total risk by accessing the value of each future by invoking \texttt{get()} on the future object. If the total risk is within the allowed risk, then the exchange reactor performs another nested asynchronous procedure call to \texttt{add\_entry} on the provider reactor with name given as a parameter to \texttt{auth\_pay}. This call results in adding an order at the target provider reactor. 

\vspace{1ex}
\minisecnodot{Does the reactor programming model necessitate manual optimization?}\\
We posit that reactors can act as a bridging model between a classic database abstraction and a key-value API. Reactors provide the possibility for developers to navigate the extremes between a single-reactor database with full SQL support and a radical decomposition of individual tuples as reactors. We envision that the typical application modeling will be a hybrid between these two extremes balancing reuse of query optimization machinery with low-level performance control. The popularity of NoSQL databases points to the need and willingness among application developers to obtain higher performance control and scalability for their applications even at the cost of sacrificing traditional database features such as query optimization and transaction support.

\subsubsection{Reactor Consistency using Transactional Semantics}
\label{sec:reactor:txn}
To guarantee consistency of the state encapsulated by a reactor database, the semantics of procedure invocations on reactors is transactional. We differentiate between top-level and nested asynchronous procedure calls. Top-level calls are executed by clients on a reactor and are termed interchangeably \emph{transactions} or \emph{root transactions}. Transactions respect the classic ACID properties~\cite{DBLP:books/aw/BernsteinHG87}. 
We denote a concrete execution $i$ of a transaction by $T_{i}$. 

Nested asynchronous procedure calls are executed by a reactor on another reactor. Since these calls must always occur within the overall context of a root transaction, they are called \emph{sub-transactions}. We denote a concrete execution of a sub-transaction $j$ of transaction $T_i$ on a reactor $k$ by $ST_{i,j}^{k}$.     
Sub-transactions allow programmers to structure their computations for performance, allowing for concurrent computation on (logically) distributed state among reactors. Sub-transactions are \emph{not} used, however, to allow for partial commitment. Any condition leading to an abort in a sub-transaction leads to the abort of the corresponding root transaction. This approach towards the semantics of nested calls is exactly the reverse of what is adopted in classic systems such as Argus~\cite{LiskovCJS87:Argus}, reflecting our focus on leveraging the high degree of \emph{physical parallelism} in modern commodity hardware for transactional processing as opposed to managing faults in settings with a high degree of physical distribution (e.g., geo-distribution) as in previous work. A transaction or sub-transaction completes only when all its nested sub-transactions complete. This frees the client logic from explicitly synchronizing on the result of a sub-transaction invocation if it does not need the result of the sub-transaction. 

For example in Figure~\ref{fig:example:app}(b), the \texttt{auth\_pay} procedure does not wait on the result of the \texttt{add\_entry} procedure call, since the programming model guarantees that the transaction corresponding to \texttt{auth\_pay} only completes when all its sub-transactions complete. 
In addition, \texttt{auth\_pay} aborts if any of the asynchronous calls to \texttt{calc\_risk} raises an abort due to excessive provider exposure.

Any program in the classic transactional model can be trivially remodeled in the reactor programming model by specifying a single reactor. For example, we could model a single exchange reactor with the schema and application logic shown in Figure~\ref{fig:example:app}(a). However, the benefits of our programming model are only fully achieved when developers of latency-sensitive OLTP applications remodel their logic as done in Figure~\ref{fig:example:app}(b). In particular, in the reformulated logic, intra-transaction parallelism is exposed. Furthermore, the trade-off between scalability on the number of provider reactors and latency of executing the logic of \texttt{auth\_pay} becomes explicit.

\vspace{1ex}
\minisecnodot{Is there a development methodology to architect an application using reactors?}\\
An interesting avenue for future research is to explore an analytical machinery for modeling and comparing the quality of reactor database designs, similar to the entity relationship model and decomposition of universal relations by functional dependencies in classic relational databases. 
Although answering this question is beyond the scope of this paper, we envision that developers could start from a single reactor type with the whole relational schema and all application functions. Through an iterative process of performance analysis and decomposition of application functionality into multiple reactor types, developers can improve latency of their applications through cross-reactor communication, and also identify inherent scalability limitations by analyzing the degree of locality in application logic.

\subsubsection{Intra-Transaction Safety}
\label{sec:reactor:safety}
Introducing asynchrony in a transactional abstraction is not trivial. Since asynchronicity exposes intra-transaction parallelism, race conditions could arise when sub-transactions that conflict on a data item are invoked asynchronously on the same reactor. Moreover, such invocations would violate the illusion
that a reactor is a computational entity with a single logical thread of control. To avoid these issues, we must enforce that at most one execution context is active for a given reactor and root transaction at any time. 

First, we enforce that whenever a reactor running a procedure directly executes a nested procedure invocation on itself, the nested invocation is executed \emph{synchronously}. This policy corresponds to inlining the sub-transaction call, resulting in future results being immediately available. 
To deal with nested asynchronous invocations, we   
define the \emph{active set} of a reactor $k$ as the set of sub-transactions, regardless of corresponding root transaction, that are currently being executed on reactor $k$, i.e., have been invoked, but have not completed. Thus, the runtime system must conservatively disallow execution of a sub-transaction $ST_{i,j}^{k}$ when: 

\vspace{-2ex}
\begin{align*}
\exists ST_{i,j'}^{k} \in \text{active\_set}(k) \; \land \; j' \neq j
\end{align*}

This dynamic safety condition prohibits programs with cyclic execution structures across reactors and programs in which different paths of asynchronous calls lead to concurrent sub-transactions on the same reactor. By introducing this safety condition, the runtime conservatively assumes that conflicts may arise in asynchronous accesses to the same reactor state within a transaction, and thus aborts any transaction with such dangerous structures. 

By leveraging this dynamic safety condition, we envision that appropriate testing of transaction logic at development time will be sufficient to root out most, if not all, dangerous structures from the code of latency-sensitive OLTP applications. However, formalizing static program checks to aid in detection of dangerous call structures among reactors is an interesting direction for future work.    

\subsection{Conflict-Serializability of Transactions}
\label{sec:math:formalism}
To formalize the correctness of concurrent executions of transactions in reactors, we show equivalence of serializable histories in the reactor model to serializable histories in the classic transactional model. We restrict ourselves exclusively to the notion of conflict-serializability. Technically, our formalization is similar to reasoning on nonlayered object transaction models~\cite{WeikumV2002:TxnBigBook}. 

\subsubsection{Background} 
We first review the formalism introduced by Bernstein et al.~\cite[page 27]{DBLP:books/aw/BernsteinHG87} for the classic transactional model and introduce relevant notation. 
In this model, the database consists of a collection of named data items, and transactions encapsulate a sequence of operations. A transaction $T_i$ is formalized as a partial ordering of operations with an ordering relation $<_{i}$ and comprises a set of operations. Operations include reads and writes, along with either a commit or an abort. A read from a data item $x$ is denoted $r_{i}[x]$, a write to $x$ denoted $w_{i}[x]$, while a commit is denoted $c_{i}$ and an abort $a_{i}$. The ordering relation $<_{i}$ orders \emph{conflicts}. Two operations conflict iff at least one of them is a write and both of them reference the same named item. We assume that a transaction does not contain multiple operations of the same type to the same named data item as in~\cite[page 27]{DBLP:books/aw/BernsteinHG87} without any impact on the~results.

\subsubsection{Reactor Model}
Without loss of generality, we assume reactor names and sub-transaction identifiers to be drawn from the set of natural numbers. Recall that we denote a sub-transaction $j$ in transaction $T_{i}$  on reactor $k$ by $ST_{i,j}^{k}$. $r_{i,j}^{k}[x]$ denotes a read from data item $x$, and $w_{i,j}^{k}[x]$ denotes a write to data item $x$ in $ST_{i,j}^{k}$. Note that data items in different reactors are disjoint. Using this notation, we can define a sub-transaction in the \textit{reactor model} as follows.

\begin{definition} \label{def:subtxn}
A sub-transaction $ST_{i,j}^{k}$ is a partial order with ordering relation
$<_{i,j}$ where,
\begin{enumerate}
\vspace{-1ex}
\item $ST_{i,j}^{k} \subseteq \{ \: r_{i,j}^{k}[x], \:
w_{i,j}^{k}[x], \: ST_{i,j'}^{k'} \: | \:$$x$ is a data item in $k$, \\$j'$ is a sub-transaction identifier s.t. $j' \neq j$ \}; 
\item Let 
\vspace{-1.5ex}
\begin{align*}
\text{basic\_ops}(r_{i,j}^{k}[x]) & = \{r_{i,j}^{k}[x]\} \\
\text{basic\_ops}(w_{i,j}^{k}[x]) & = \{w_{i,j}^{k}[x]\} \\
\text{basic\_ops}(ST_{i,j}^{k}) & = \{ \text{basic\_ops(o)} \; | \; o \in ST_{i,j}^{k} \},
\end{align*}
\vspace{-0.5ex}
if $o_1 \in ST_{i,j}^{k} \wedge o_2 \in ST_{i,j}^{k}$ \\
  $ \text{                         } \wedge r_{i,j'}^{k'}[x] \in \text{basic\_ops}(o_1)  $ \\ 
  $ \text{                         } \wedge w_{i,j''}^{k'}[x] \in \text{basic\_ops}(o_2)$,\\ 
  then either $o_1 <_{i,j} o_2$ or $o_2 <_{i,j} o_1$. 
\end{enumerate}
\vspace{-1ex}
\end{definition}

Note that the ordering relation $<_{i,j}$ of a sub-transaction establishes order according to conflicts in leaf-level basic operations, potentially nested in sub-transactions. These leaf-level basic operations are determined by the function basic\_ops in the definition.

\begin{definition}
A transaction $T_{i}$ is a partial order with ordering relation
$<_{i}$ where,
\begin{enumerate}
\vspace{-1ex}
\item $T_{i} \subseteq \{ \: ST_{i,j}^{k} \: \} \cup \{ a_{i},
c_{i} \};$
\item $a_{i} \in T_{i} \iff c_{i} \not \in T_{i};$
\item $\text{if t is} \; c_{i} \; or \; a_{i} \;(\text{whichever
is in} \; T_{i}), \text{then for every other}\\\text{operation } p \in
T_{i}, p <_{i} t; $
\item if $o_1 \in T_{i} \wedge o_2 \in T_{i} \wedge o_1,o_2 \not \in \{ a_{i}, c_{i} \}  \wedge$ \\ 
 $ \text{                         } r_{i,j'}^{k'}[x] \in \text{basic\_ops}(o_1)  \wedge w_{i,j''}^{k'}[x] \in \text{basic\_ops}(o_2)$,\\ 
  then either $o_1 <_{i} o_2$ or $o_2 <_{i} o_1$. 
   
\end{enumerate}
\vspace{-1ex}
\end{definition}

Formally, a transaction comprises exclusively sub-transactions, and the relation $<_{i}$ orders sub-transactions according to conflicts in their nested basic operations. In the reactor model, two sub-transactions \emph{conflict} iff the basic operations of at least one of them contain a write and the basic operations of both of them reference the same named item in the same reactor. Under this extended notion of a conflict, the definition of history, serial history, equivalence of histories and serializable history in the reactor model are the same as their definitions in the classic transactional model~\cite{DBLP:books/aw/BernsteinHG87}, but with sub-transactions replacing basic operations. Similar to nested transaction models~\cite{Beeri:1989:JACM}, we then wish to establish an equivalence between serializability of transactions in the reactor model and serializability in the classic transactional model. To do so, 
we proceed by defining an appropriate projection of the reator model into the classic transactional model. 

\begin{definition}
\label{proj-tim-tmm-simple-op}
The projection of a basic operation $o$ from the reactor model to the classic transactional model, denoted by $P(o)$, is defined as:
\begin{enumerate}
\item $P(r_{i,j}^{k}[x]) = r_{i}[k \circ x]$  
\item $P(w_{i,j}^{k}[x]) = w_{i}[k \circ x]$
\item $P(c_{i}) = c_{i}$
\item $P(a_{i}) = a_{i}$
\end{enumerate}
where $\circ\text{ denotes concatenation}$.
\end{definition}

The definition provides a name mapping from the disjoint address spaces of reactors
to a single address space, which is done by concatenating the reactor identifier with name for a data item.

\begin{definition}
\label{proj-tim-tmm-subtxn}
The projection of a sub-transaction $ST_{i,j}^{k}$ from the reactor model to the classic transactional model, denoted by $P_{S}(ST_{i,j}^{k})$ is a partial order with ordering relation $<_{S}^{i,j}$: 
\begin{enumerate}
\item  $P_{S}(ST_{i,j}^{k}) \subseteq  \{ P(o) \; | \; o \in \text{basic\_ops}(ST_{i,j}^{k})$\};
\item if $o_1 \in ST_{i,j}^{k} \wedge o_2 \in ST_{i,j}^{k} \wedge \; o_1 <_{i,j} o_2 \; \wedge \; o_1,o_2 \text{ are reads or writes}$, then $P(o_1) <_{S}^{i,j} P(o_2)$;
\item if $ST_{i,j'}^{k'} \in ST_{i,j}^{k}$, then $<_{S}^{i,j}$ is extended by  $<_{S}^{i,j'}$;
\item if $o_1 \in ST_{i,j}^{k} \wedge o_1 \text{ is a read or a write} \wedge ST_{i,j'}^{k'} \in ST_{i,j}^{k} \wedge \; o_1 <_{i,j} ST_{i,j'}^{k'}$, then $P(o_1) <_{S}^{i,j} P_{S}(ST_{i,j'}^{k'})$; 
\item if $ST_{i,j'}^{k'} \in ST_{i,j}^{k} \wedge o_2 \in ST_{i,j}^{k} \wedge o_2 \text{ is a read or a write} \wedge ST_{i,j'}^{k'} <_{i,j} o_2$, then $P_{S}(ST_{i,j'}^{k'}) <_{S}^{i,j} P(o_2)$;
\item if $ST_{i,j'}^{k'} \in ST_{i,j}^{k} \wedge ST_{i,j''}^{k''} \in ST_{i,j}^{k} \wedge ST_{i,j'}^{k'} <_{i,j} ST_{i,j''}^{k''}$, then $P_{S}(ST_{i,j'}^{k'}) <_{S}^{i,j} P_{S}(ST_{i,j''}^{k''})$.
\end{enumerate}
\end{definition}
 
\begin{definition}
\label{proj-tim-tmm}
The projection of a transaction $T_i$ from the reactor model to the classic transactional model, denoted by $P_{T}(T_{i})$ is a partial order with ordering relation $<_{T}^{i}$:  
\begin{enumerate}
\item $P_{T}(T_{i}) \subseteq ( \bigcup_{ST_{i,j}^{k} \in T_i}  P_{S}(ST_{i,j}^{k})) \; \cup  \; \{P(o) \; | \;\ o \in T_i \; \wedge o \; \text{ is a commit or abort}\}$;
\item $<_{T}^{i}$ is extended by $\bigcup_{ST_{i,j}^{k} \in T_i}<_{S}^{i,j}$;
\item if $ST_{i,j}^{k} \in T_i \wedge ST_{i,j'}^{k'} \in T_i \wedge ST_{i,j}^{k} <_{i} ST_{i,j'}^{k'} \wedge o_1 \in P_S(ST_{i,j}^{k}) \wedge o_2 \in P_S(ST_{i,j'}^{k'})$, then $o_1 <_{T}^{i} o_2$;
\item $\text{if t is} \; c_{i} \; or \; a_{i} \;(\text{whichever
is in} \; P_{T}(T_{i})), \text{for any further}\\ \text{operation} \; p \in
P_{T}(T_{i}), p <_{T}^{i} t$.
\end{enumerate}
\end{definition}

The definitions unroll all sub-transactions in the reactor model into read and write operations in the classic transactional model while maintaining ordering constraints. 

\begin{definition}
\label{history-tim}
The projection $P(H)$ of a history H over a set of transactions T = $\{T_{1},T_{2}, ..., T_{n}\}$ from the reactor model to the classic transactional model is a partial order with ordering relation $<_{P_H}$ over a set of transactions $T'$ = $\{P_{T}(T_{1}),P_{T}(T_{2}), ..., P_{T}(T_{n})\}$ iff:
\begin{enumerate}
\item $P_{P_H}(H) = \bigcup_{i=1}^{n}P_{T}(T_{i});$
\item $<_{P_H}$ is extended by $\bigcup_{i=1}^{n}  <_{T}^{i};$
\item if $o_1 \in P_S(ST_{i,j}^{k}) \wedge o_2 \in P_S(ST_{i',j'}^{k'}) \wedge ST_{i,j}^{k} \in T_{i} \wedge ST_{i',j'}^{k'} \in T_{i'} \wedge ST_{i,j}^{k} <_{H} ST_{i',j'}^{k'}$, then $o_1 <_{P_H} o_2$ as long as $o_1$ and $o_2$ conflict.
\end{enumerate}
\end{definition}

\begin{theorem}\label{THM:EQUIVALENCE}
A history $H$ is serializable in the reactor model iff its projection $H'=P(H)$ in the classic transactional model is serializable. 
\end{theorem}
\vspace{-1.5ex}
\begin{proof}
See Appendix~\ref{thm:equivalence:proof}.
\end{proof} 
\vspace{-1ex}

Theorem~\ref{THM:EQUIVALENCE} implies that, with appropriate care, a scheduler for the classic transactional model can be used to implement one for the reactor model. In Section~\ref{sec:sys:arch}, we reuse an optimistic concurrency control (OCC) scheduler~\cite{Tu:2013:STM:2517349.2522713} and two-phase commit (2PC)~\cite{DBLP:books/aw/BernsteinHG87}.

\subsection{Computational Cost Model}
\label{sec:perf:predict}
In this section, we introduce a cost model to support developers of latency-sensitive applications in controlling the latency of a transaction program expressed using reactors. Clearly, latency depends heavily on program structure. For example, certain programs can overlap asynchronous invocations of functions in other reactors with processing logic; 
other programs may do so only conditionally, or have data dependencies between different asynchronous function calls. For concreteness, we focus on a subset of programs modeled after a fork-join parallelism pattern~\cite{LeijenSB09:TPL}. 

\begin{figure}[!t]
\begin{align*}
L&(ST_{i,j}^{k}) =  P_{seq}(ST_{i,j}^{k}) \quad + \sum_{ST_{i,j'}^{k'} \in \; \text{sync}_{seq}(ST_{i,j}^{k})} L(ST_{i,j'}^{k'}) \quad  \\
& + \quad  \sum_{k' \in \;\text{dest}\big(\text{sync}_{seq}(ST_{i,j}^{k})\big)} \big(C_s(k,k') \;+ C_r(k',k)\big) \quad \\
&
+ \hspace{1.5ex} max\Bigg(max_{ST_{i,j'}^{k'} \in \; \text{async}(ST_{i,j}^{k})} \bigg(L(ST_{i,j'}^{k'})+C_r(k',k) \\
& \hspace{11ex} + \sum_{k'' \in \;\text{dest}\big(\text{prefix}(\text{async}(ST_{i,j}^{k}) \;,\;ST_{i,j'}^{k'})\big)}C_s(k,k'')\bigg), \\
&\hspace{6ex}P_{ovp}(ST_{i,j}^{k}) \quad + \sum_{ST_{i,j'}^{k'} \in \; \text{sync}_{ovp}(ST_{i,j}^{k})} L(ST_{i,j'}^{k'}) \quad + \\ 
& \quad \sum_{k' \in \; \text{dest}\big(\text{sync}_{ovp}(ST_{i,j}^{k})\big)}\big(C_s(k,k') + C_r(k',k)\big)\Bigg)
\end{align*}
\vspace{-3ex}
\caption{Modeling latency cost of a fork-join sub-transaction in the reactor model.}
\label{fig:cost:model}
\vspace{-4ex}
\end{figure}

\minisec{Fork-Join Sub-Transactions}
We call a sub-transaction fork-join if it comprises: (a) sequential logic, potentially involving synchronous calls to sub-transactions; (b) parallel logic consisting of calls to children fork-join sub-transactions such that all asynchronous invocations happen simultaneously at one given program point only, are potentially overlapped with synchronous logic, and then all future results are collected. Consider one such sub-transaction $ST_{i,j}^{k}$. We call $\text{sync}_{seq}(ST_{i,j}^{k})$ its sequence of children sub-transactions and $P_{seq}(ST_{i,j}^k)$ its processing logic executed sequentially. 
To capture communication costs, 
we term $C_s(k,k')$ the cost to send a sub-transaction call from reactor $k$ to reactor $k'$, and $C_r(k',k)$ the cost to receive a result from $k'$ at $k$. The sequence of children sub-transactions of $ST_{i,j}^{k}$ executed asynchronously are denoted $\text{async}(ST_{i,j}^{k})$. The synchronous children sub-transactions and processing logic overlapped with the asynchronous sub-transactions are represented by $\text{sync}_{ovp}(ST_{i,j}^{k})$ and $P_{ovp}(ST_{i,j}^k)$, respectively. Given a sequence $S$ of sub-transactions, $\text{prefix}(S,ST_{i,j}^k)$ denotes the sequence of sub-transactions in $S$ up to $ST_{i,j}^k$ and including it. Moreover, we say that $\text{dest}(ST_1, \ldots ,ST_n)$ represents the sequence of reactors that sub-transactions $ST_1, \ldots ,ST_n$ execute on.

Now, the latency cost of $ST_{i,j}^{k}$ is modeled by the formula in Figure~\ref{fig:cost:model}, assuming the parallelism in asynchronous sub-transactions is fully realized. 
The same formula can be applied recursively to compute the latency cost for sub-transactions of arbitrary depth. Since a root transaction is a special case of a sub-transaction, i.e., a sub-transaction without a parent, the same formula applies, modulo any overheads incurred for commitment.

\minisec{Uses}
Many applications can be written in the reactor model with fork-join sub-transactions. Consider the transaction \texttt{auth\_pay} in Figure~\ref{fig:example:app}(b). While it is not fork-join as presented, it can be easily decomposed into two sequentially invoked fork-join sub-transactions, namely one to calculate \texttt{total\_risk} and one to conditionally call \texttt{add\_entry}. All the benchmarks evaluated in our experiments can also be expressed with fork-join sub-transactions. Notwithstanding, the reactor programming model allows for any synchronization pattern with futures respecting the conditions of Section~\ref{sec:reactor:safety}, and is \emph{not limited} to fork-join sub-transactions. The cost model of Figure~\ref{fig:cost:model} can be extended to cover other program classes, if necessary.

We envision that developers may employ cost modeling as in Figure~\ref{fig:cost:model} in a way similar to algorithmic complexity measures, e.g.,~\cite{AggarwalV88:IOModel}, to compare alternative formulations for transaction programs. 
Developers can improve the latency of their sub-transactions by: (1)~increasing asynchronicity of children sub-transactions, (2)~overlapping execution of application logic by introducing sub-transactions, and (3)~reducing the processing cost of the application logic.

Finally, developers can reason about scalability by considering how data, computation and communication are spread among reactors. In particular, if developers architect their applications such that increasing amounts of data and computation are distributed among increasing numbers of reactors while at the same time keeping the number of cross-reactor calls roughly constant per transaction, then adequate transactional scalability should be expected.

\minisec{Limitations}
Again similar to algorithmic complexity measures, a cost model such as the one in Figure~\ref{fig:cost:model} is exposed to limitations in the estimation of its various parameters for concrete fork-join transaction programs and system realizations of the reactor model. It may be impossible to estimate reliably how costly processing will be or how many sub-transaction invocations will fall into $\text{sync}_{seq}(ST_{i,j}^{k})$, $\text{sync}_{ovp}(ST_{i,j}^{k})$, and $\text{async}(ST_{i,j}^{k})$, as these parameters may be determined by complex program logic and data dependencies. In addition, the concrete system realization may not express the full parallelism encoded in the program. Moreover, determination of cost model parameters such as $C_s$ and $C_r$ may be compromised by measurement errors. Finally, the cost model considers transaction programs in isolation, and does not include interference or queueing effects. Notwithstanding, we remark in our experimental evaluation that under certain conditions, the cost model can closely capture latency differences between alternative transaction programs.

\section{System Architecture}
\label{sec:sys:arch}

\techreport{In this section, we discuss the architecture of \reactdb, an in-memory database system that exposes the reactor programming model. There are many challenges that must be met by \reactdb's architecture. First, \reactdb\ must honor the high-level cost model for reactors as discussed in Section~\ref{sec:perf:predict}. Second, \reactdb\ must provide for transactional guarantees, leveraging the results of Section~\ref{sec:math:formalism}. Third, \reactdb\ must provide low-level control to infrastructure engineers of overhead and program-to-data affinity to maximize efficiency. We describe how \reactdb\ addresses the first two challenges in Sections~\ref{sec:reactdb:overview} and~\ref{sec:reactdb:cc}, and the third challenge in Section~\ref{sec:deployments}.}

\subsection{Overview}
\label{sec:reactdb:overview}
In this section, we discuss the architecture of \reactdb, an in-memory database system that exposes the reactor programming model.
The design of \reactdb\ aims at providing control over the mapping of reactors to physical computational resources and memory regions under concurrency control. The system implementation currently targets a single multi-core machine for deployment; however, \reactdb's architecture is designed to allow for deployments in a cluster of machines, which we leave for future work.

\label{sec:architecture}
\begin{figure}[!t]
\centering
    \scalebox{0.28}[0.27]{\includegraphics{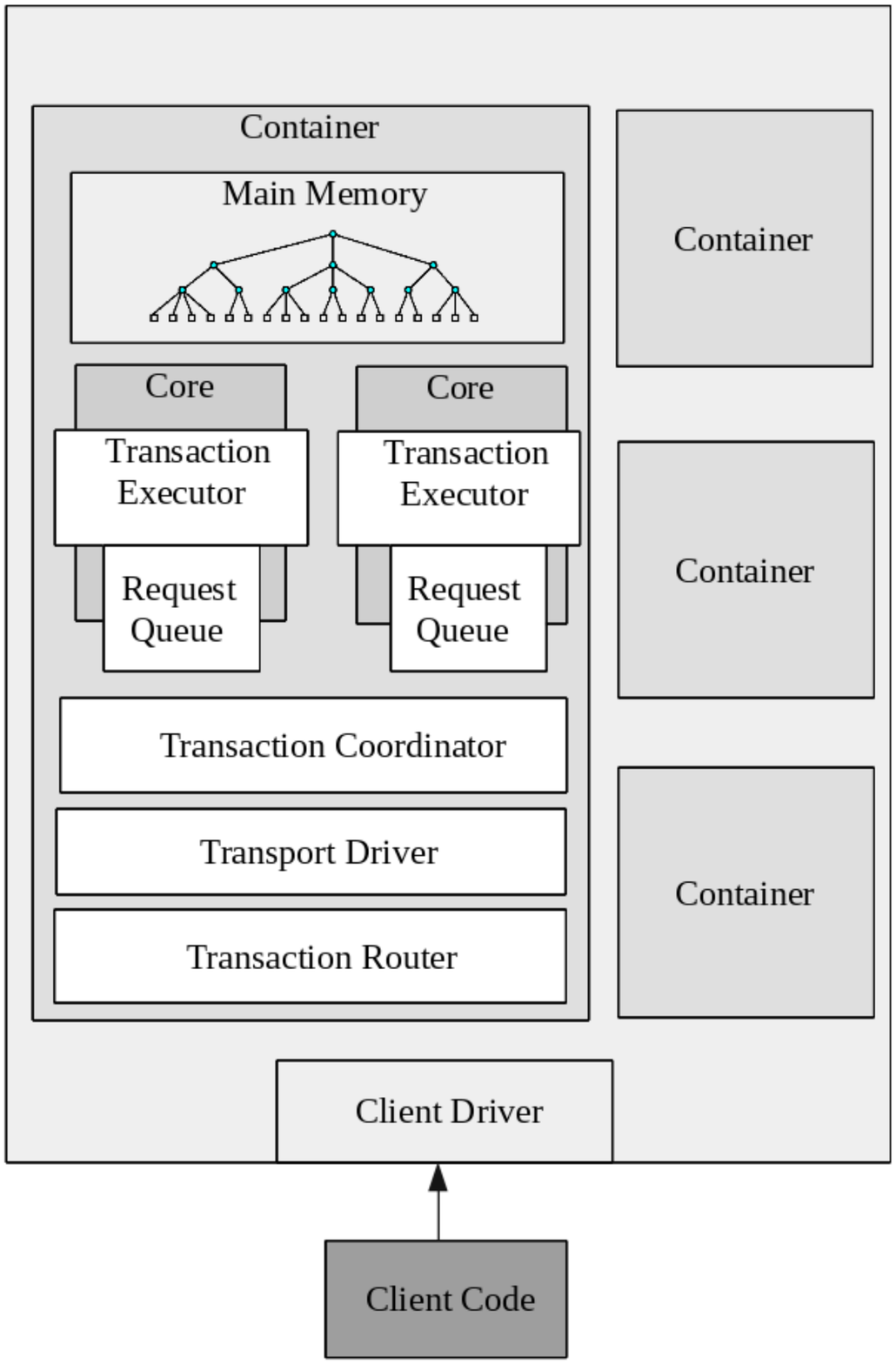}} 
    \vspace{-2ex}
  \caption{\reactdb's architecture.}
    \vspace{-2ex}
  \label{arch:reactdb}
\end{figure}

As shown in Figure~\ref{arch:reactdb}, \reactdb's architecture is organized as a collection of database \emph{containers}. A container abstracts a (portion of a) machine with its own storage (main memory) and associated mechanisms for transactional consistency. Each container is isolated and does not share its data with other containers. Containers are associated with computational resources (cores) disjoint from other containers, abstracted by \emph{transaction executors}. A transaction executor consists of a thread pool and a request queue, and is responsible for executing requests, namely asynchronous procedure calls. Each transaction executor is pinned to a core. Single-container transactions are managed by the concurrency control mechanism within the container, while a \emph{transaction coordinator} runs a commitment protocol for transactions spanning multiple containers. Transactional durability is currently unavailable in our implementation, but could be achieved by a combination of techniques such as fast log-based recovery~\cite{ZhengTKL14:SiloR} and distributed checkpoints~\cite{ElnozahyAWJ02:DistCkpt}. Alternatively, an approach such as FaRM's could be employed to minimize any impact of durability mechanisms on latency~\cite{Dragojevic:2015:FaRM}. 

Each container stores a two-level mapping between a reactor and a transaction executor. On the first level, a reactor is mapped to one and only one container. Together with appropriate container deployment, this constraint ensures that asymmetrically large communication costs are only introduced between, but not within, reactors, 
in line with our computational cost model. 
On the second level, a reactor can be mapped to one or more transaction executors in a container. \emph{Transaction routers} decide the transaction executor that should run a transaction or sub-transaction according to a given policy, e.g., round-robin or affinity-based. 

\emph{Transport drivers} handle communication across containers. \reactdb\ has a driver component that is used by client code to send transactions into the system for processing. \reactdb\ accepts pre-compiled stored procedures written in the reactor programming model in C++ against a record manager interface. An instance of a pre-compiled stored procedure and its inputs forms a transaction. 

\subsection{Concurrency Control}
\label{sec:reactdb:cc}
\subsubsection{Single Container Transactions}
Every transaction or sub-transaction written in the reactor programming model specifies the reactor where it must be executed. 
If the destination reactor of a child sub-transaction is hosted in the same container as the parent sub-transaction, the child sub-transaction is executed synchronously within the same transaction executor to minimize the communication overhead of migrating across transaction executors. If all the sub-transactions in the execution context of a root transaction are executed within one container, then the native concurrency control mechanism of the container is used to guarantee serializability. As a consequence of Theorem~\ref{THM:EQUIVALENCE}, \reactdb\ can reuse an existing concurrency control mechanism, and we chose Silo's high-performance OCC implementation~\cite{Tu:2013:STM:2517349.2522713}. 

\subsubsection{Multi-Container Transactions}
When a sub-transaction is invoked on a reactor mapped to a container different than the current container, the call is routed by the transport driver to the destination container and then by the transaction router to the request queue of a transaction executor.  
Once the sub-transaction is queued, the calling code gets a future back representing this computation. If the calling sub-transaction code does not synchronize on the future, then once the caller completes, \reactdb\ enforces synchronization on the futures of all child sub-transactions. 
By the above synchronization policy, a root transaction can finish when all the sub-transactions created and invoked in its context finish, recursively. 
The transaction executor then invokes the transaction coordinator to initiate a commitment protocol across the containers that have been touched by the transaction, either directly or by any of its nested sub-transactions. The transaction coordinator in turn performs a 2PC protocol. The first phase of the protocol triggers validation of Silo's OCC protocol on all the involved containers, during which locks are acquired on the write-set of the transaction. If any of the validations fail, the second phase of the protocol ensures that the transaction is aborted on all containers. Otherwise, the write phase of Silo's OCC scheme is triggered. In either case, all locks are released appropriately~\cite{Tu:2013:STM:2517349.2522713}.

\subsubsection{Thread Management}
To minimize the effect of stalls due to synchronization, each transaction executor maintains a thread pool to process (sub-)transactions. A configurable number of threads are allowed to become active and drain the transactor executor's request queue, thus controlling the multi-programming level (MPL) per executor. In addition, the threads use cooperative multitasking to minimize context switching overheads. A thread blocks if it tries to access the result of a sub-transaction invoked on a different container and the result is not yet available. In such a situation, it notifies another thread to take over processing of the request queue and goes back to the thread pool when the (sub-)transaction being executed by it is completed. 

\subsection{Deployments}
\label{sec:deployments}
Configuration of transaction executors and containers allows infrastructure engineers to flexibly deploy \reactdb\ in a number of database architectures. 
In the remainder of the paper, we restrict ourselves to three main deployment strategies: 

\noindent\textbf{(S1)} \textsf{shared-everything-without-affinity}: This strategy employs a single container in which each transaction executor can handle transactions on behalf of any reactor. \reactdb\ is configured with a round-robin router to load balance transactions among executors. All sub-transactions are executed within the same transaction executor to avoid any migration of control overhead. This strategy adheres to the architecture of most shared-everything databases~\cite{HellersteinSH07:ArchDB}.

\noindent\textbf{(S2)} \textsf{shared-everything-with-affinity}: This strategy is similar to \textsf{shared-everything-without-affinity} in that it employs a single container, but with the difference that an affinity-based router ensures that root transactions for a given reactor are processed by the same transaction executor.  
In sub-transaction calls, even if to different reactors, no migration of control happens, and the sub-transaction is executed by the same transaction executor of the root transaction. This deployment strategy closely adheres to the setup employed in the evaluation of Silo~\cite{Tu:2013:STM:2517349.2522713}. 

\noindent\textbf{(S3)} \textsf{shared-nothing}: This strategy employs as many containers as transaction executors, and a given reactor is mapped to exactly one transaction executor. While this strategy aims at maximizing program-to-data affinity, sub-transaction calls to different reactors may imply migration of control overheads to other transaction executors. We further decompose this configuration into \textsf{shared-nothing-sync} and \textsf{shared-nothing-async},  depending on how sub-transactions are invoked within application programs. In the former option, sub-transactions are invoked synchronously by calling \texttt{get} on the sub-transaction's future immediately after invocation. In the latter, the call to \texttt{get} is delayed as much as possible for maximal overlapping of application logic with sub-transaction calls. From an architecture perspective, both of these setups represent a \textsf{shared-nothing} deployment with differing application programs exercising different synchronization options. The \textsf{shared-nothing-sync} strategy models the setup of shared-nothing databases such as H-Store~\cite{Stonebraker:2007:EAE:1325851.1325981} and HyPer~\cite{Kemper:2011:HHO:2004686.2005619}, albeit with a different concurrency control protocol due to potentially higher MPL per executor. The \textsf{shared-nothing-async} strategy allows \reactdb\ to further leverage intra-transaction parallelism as provided by the reactor programming model, exploiting cooperative multitasking.

Other flexible deployments, similar to~\cite{PorobicPBTA12:HardwareIslands}, are possible as well. To change database architecture, only configuration files need to be edited and the system bootstrapped, without changes to application logic operating on reactors.

\section{Evaluation}
\label{sec:evaluation}
In this section, we evaluate the effectiveness of \reactdb\ and the reactor programming model. The experiments broadly aim at validating the following hypotheses: 
\noindent \textbf{(H1)}~The reactor programming model allows for reasoning about latency in alternative formulations of application programs (Section~\ref{sec:exp:pred:programs} and Appendix~\ref{sec:exp:pred:cost}).
\noindent \textbf{(H2)}~The computational cost model of reactors can be efficiently realized by \reactdb, subject to its limitations (Section~\ref{sec:exp:pred:programs:breakdown} and Appendix~\ref{sec:exp:pred:limitations}).
\noindent \textbf{(H3)}~\reactdb\ allows for configuration of database architecture, without any changes to application code, so as to exploit asynchronicity in transactions depending on the level of load imposed on the database (Section~\ref{sec:exp:virt} and Appendices~\ref{sec:cost:new:order} and~\ref{sec:cross:reactor}).
We present additional evidence of \reactdb's transactional scale-up capability in Appendix~\ref{sec:scaleup:overhead}. Furthermore, we evaluate procedure-level parallelism with reactors based on the scenario of Figure~\ref{fig:example:app} in Appendix~\ref{sec:procedure:parallelism}.

\subsection{Experimental Setup}
\label{sec:exp:setup}
\subsubsection{Hardware}
For our latency measurements in Section~\ref{sec:exp:pred} and Appendices~\ref{sec:exp:pred:cost} and~\ref{sec:exp:pred:limitations}, we employ a machine with one four-core, 3.6 GHz Intel Xeon E3-1276 processor with hyperthreading, leading to a total of eight hardware threads. Each physical core has a private 32 KB L1 cache and a private 256 KB L2 cache. All the cores share a last-level L3 cache of 8 MB. The machine has 32 GB of RAM and runs 64-bit Linux 4.1.2. A machine with high clock frequency and uniform memory access was chosen for these experiments to challenge our system's ability to reflect low-level latency asymmetries in modern hardware as captured by our programming model. 

For Section~\ref{sec:exp:virt} and Appendices~\ref{sec:cost:new:order}--\ref{sec:procedure:parallelism}, we use a machine with two sockets, each with 8-core 2.1 GHz AMD Opteron 6274 processors including two physical threads per core, leading to a total of 32 hardware threads. Each physical thread has a private 16 KB L1 data cache. Each physical core has a private 64 KB L1 instruction cache and a 2 MB L2 cache. Each of the two sockets has a 6 MB L3 cache. The machine has 125 GB of RAM in total, with half the memory attached to each of the two sockets, and runs 64-bit Linux 4.1.15. The higher number of hardware threads and accentuated cache coherence and cross-core synchronization effects allow us to demonstrate the effect of virtualization of database architecture in experiments varying transaction load and asynchrony. 

\subsubsection{Methodology}
An epoch-based measurement approach similar to Oltpbench is used~\cite{DBLP:journals/pvldb/DifallahPCC13}.~Average latency or throughput is calculated across 50 epochs and the standard deviation is plotted in error bars.  All measurements include the time to generate transaction~inputs. 

\subsubsection{Workloads and Deployments}
For the experiments of Section~\ref{sec:exp:pred}, we implement an extended version of the Smallbank benchmark mix~\cite{DBLP:conf/icde/AlomariCFR08}. Smallbank simulates a banking application where customers access their savings and checking accounts. 
Oltpbench first extended this benchmark with a transfer transaction, which is implemented by a credit to a destination account and a debit from a source account~\cite{hstore-benchmarks}. We extend the benchmark further with a multi-transfer transaction. Multi-transfer simulates a group-based transfer, i.e., multiple transfers from the same source to multiple destinations. Thus, by varying the number of destination accounts for multi-transfer and controlling the deployment of \reactdb, we can vary both the amount of processing in the transaction as well as the amount of cross-reactor accesses that the transaction makes. 

Each customer is modeled as a reactor. We configure \reactdb\ with 7 database containers, each hosting a single transaction executor for a total of 7 transaction executors mapped to 7 hardware threads. The deployment plan of \reactdb\ is configured so that each container holds a range of 1000 reactors. A single worker thread is employed to eliminate interference effects and allow us to measure latency overheads of single transactions. The worker thread generating transaction inputs and invocations is allocated in a separate worker container and pinned to the same physical core hosting the container responsible for the first range, but in a separate hardware thread. In order to keep our measurements comparable, the multi-transfer transaction input generator always chooses a source customer account from this first container. 

The experiments of Section~\ref{sec:exp:virt} use the classic TPC-C benchmark~\cite{TPCC}. We closely follow the implementation of the benchmark from Oltpbench~\cite{hstore-benchmarks}, which makes the usual simplifications, e.g., regarding think times. 
In our port of TPC-C, we model each warehouse as a reactor, and configure database containers differently according to the experiment. We vary the number of client worker threads generating transaction invocations, and group these workers into a worker container separate from the database containers that host transaction executors and carry out transaction logic.  
Each client worker thread generates load for only one warehouse (reactor), thus modeling client affinity to a warehouse. To showcase \reactdb's ability to configure database architecture at deployment time, we experiment with the deployments described in Section~\ref{sec:deployments}.

\subsubsection{Application Programs}
We evaluate different application program formulations for the multi-transfer transaction added to Smallbank, exercising the asynchronous programming features of reactors.
Similar to Figure~\ref{fig:example:app}(b), multi-transfer invokes multiple sub-transactions. In contrast to the figure, in some program variants, we force synchronous execution by immediately calling \texttt{get} on the future returned. The first formulation, \textsf{fully-sync}, invokes multiple transfer sub-transactions from the same source synchronously. Each transfer sub-transaction in turn invokes a synchronous credit sub-transaction on the destination account and a synchronous debit sub-transaction on the source account. The \textsf{partially-async} formulation behaves similarly; however, each transfer sub-transaction invokes an asynchronous credit on the destination account and a synchronous debit on the source account, overlapping half of the writes in the processing logic while still executing communication proportional to the transaction size sequentially. The \textsf{fully-async} formulation does not invoke transfer sub-transactions, but rather explicitly invokes asynchronous credit sub-transactions on the destination accounts and multiple synchronous debit sub-transaction on the source account. Thus, not only are roughly half of the writes overlapped, but also a substantial part of the communication across reactors. The final formulation, \textsf{opt}, is similar to the \textsf{fully-async} transaction, but performs a single synchronous debit to the source account for the full amount instead of multiple debits. Consequently, processing depth is further reduced and should roughly equal two writes, while communication should be largely overlapped. The implementation of the above program formulations using the reactor programming model is available in Appendix \ref{sec:smallbank:reactor:implementation}. 

In addition, we implement all transactions of TPC-C in our programming model. Unless otherwise stated, we overlap calls between reactors as much as possible in transaction logic by invoking sub-transactions across reactors asynchronously. 

\begin{figure*}[!t]
        \begin{minipage}{0.32\linewidth}
        \centerline{\includegraphics[width=0.95\linewidth]{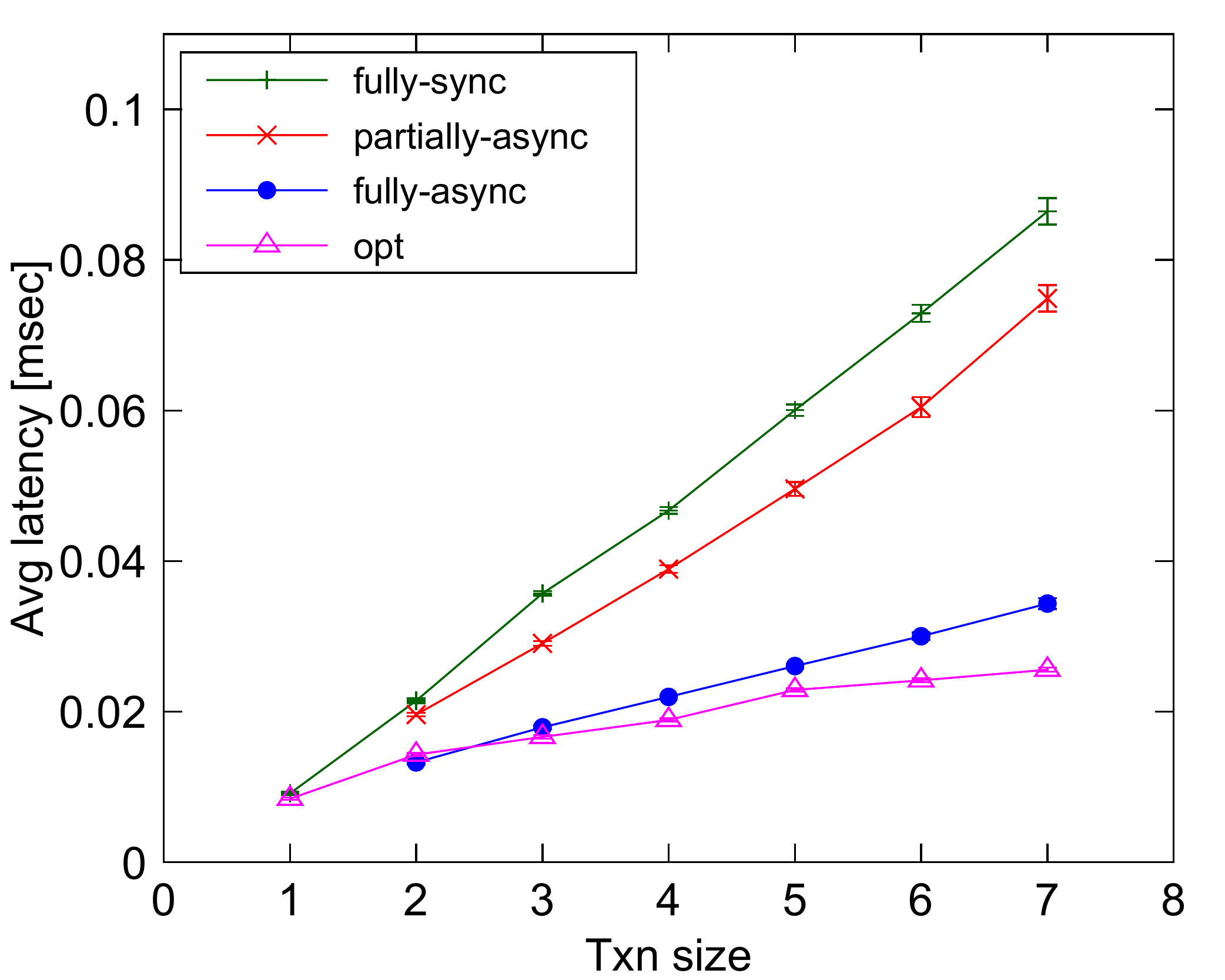}}
        \vspace{-1ex}
        \caption{Latency vs. size and different user program formulations.} \label{fig:latency:programs}
    \end{minipage} \hfill
    \begin{minipage}{0.32\linewidth}
        \centerline{\includegraphics[width=0.95\linewidth]{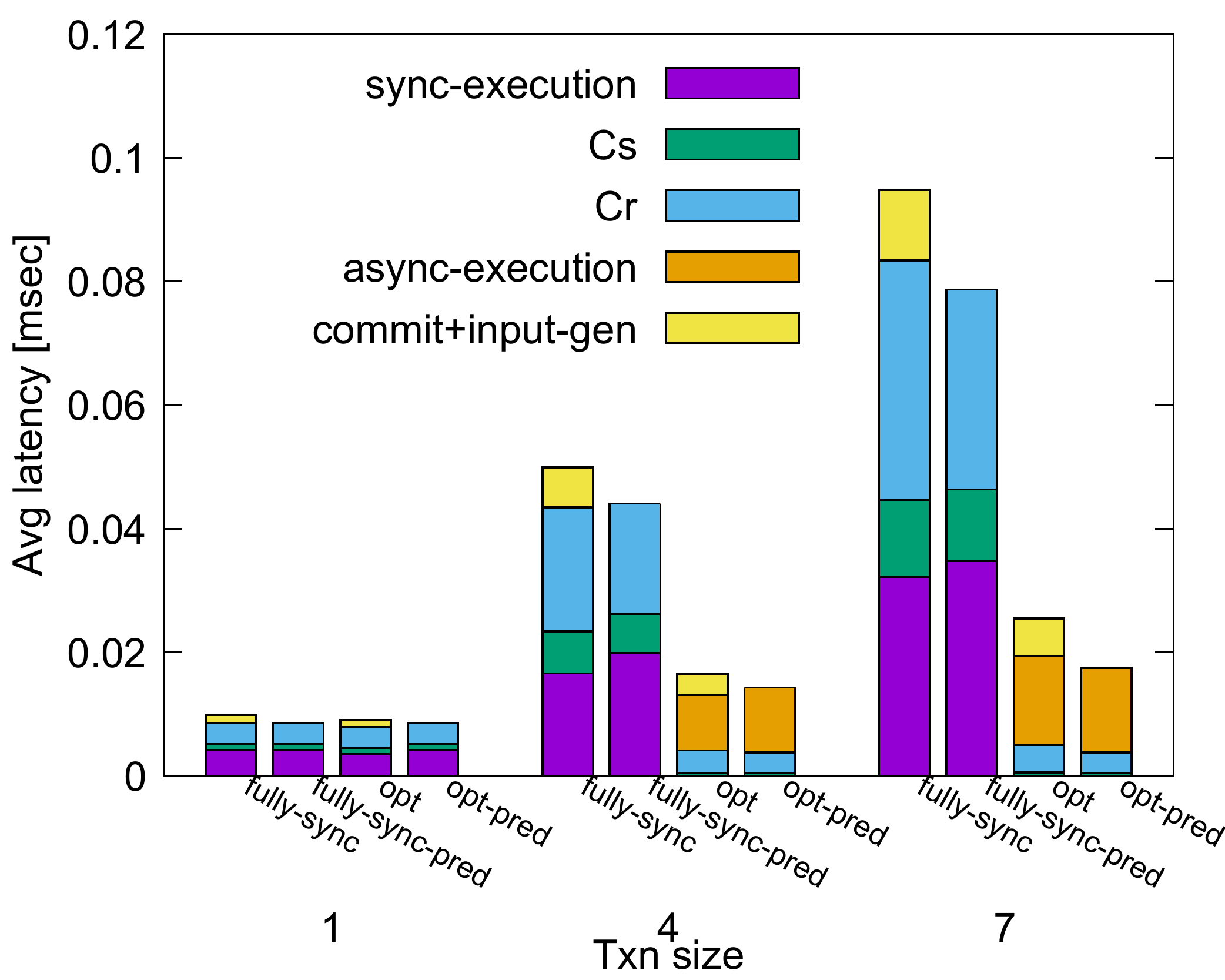}}
        \vspace{-1ex}
        \caption{Latency breakdown into cost model components.} \label{fig:latency:cost:model}
    \end{minipage} \hfill
    \begin{minipage}{0.32\linewidth}
      \centerline{\includegraphics[width=0.95\linewidth]{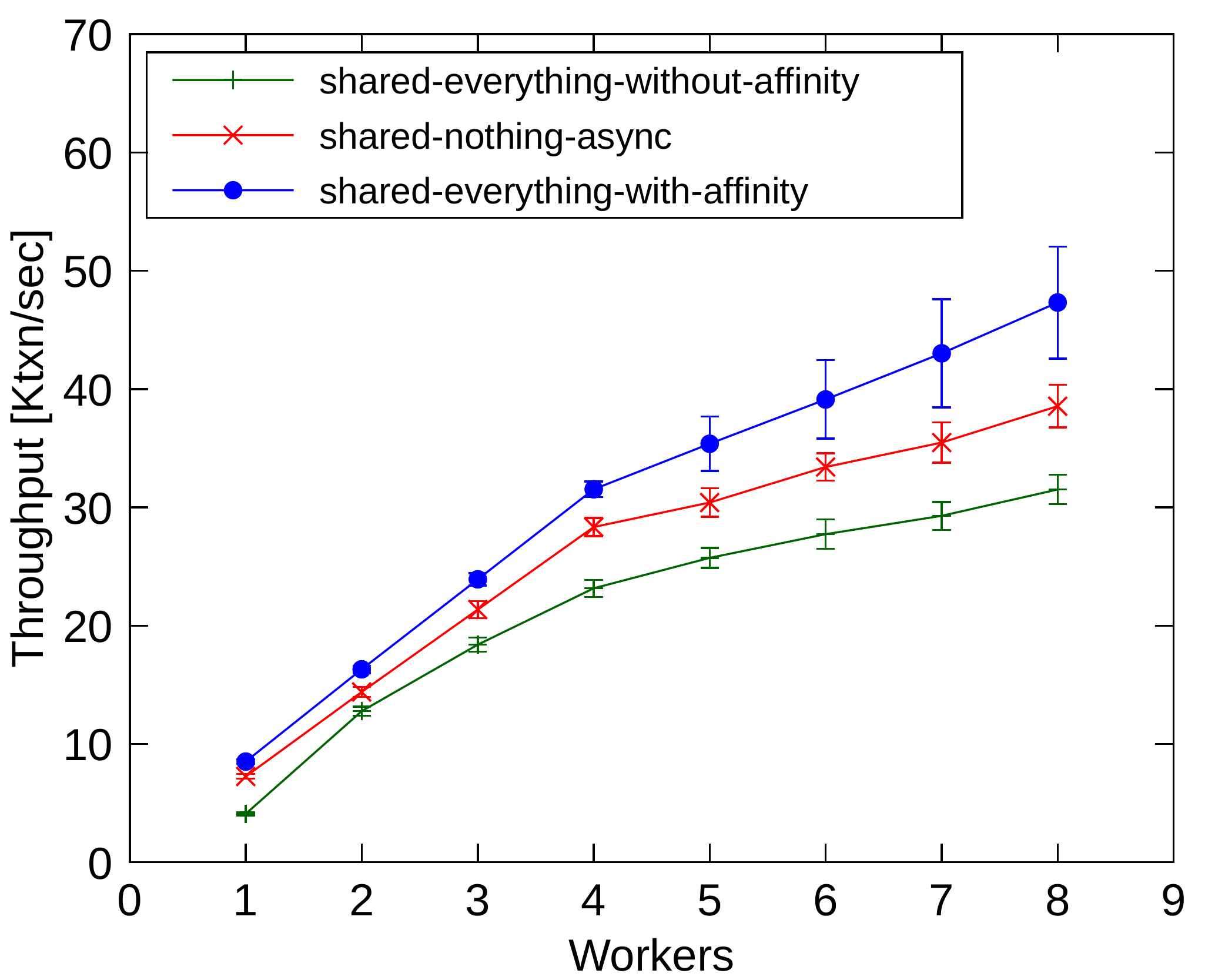}}
      \vspace{-1ex}
      \caption{TPC-C throughput with varying load at scale factor 4.}\label{fig:throughput:fixed-db-load:deployments}
    \end{minipage} \hfill
\vspace{-2ex}
\end{figure*}

\subsection{Latency Control}
\label{sec:exp:pred}
\subsubsection{Latency vs. Program Formulations}
\label{sec:exp:pred:programs}
In this section, we show an experiment in which 
we vary the size of a multi-transfer transaction by increasing the number of destination accounts. Each destination is chosen on a different container out of the seven in our shared-nothing deployment. 
The latency for the different application program formulations is outlined in Figure \ref{fig:latency:programs}.~The observed curves match the trends predicted by the cost equation of Figure~\ref{fig:cost:model}. First, as we increase transaction size, the processing and communication costs of a multi-transfer increase linearly across all formulations. Second, the highest latencies overall are for \textsf{fully-sync}, and latencies become lower as more asynchronicity is introduced in the formulations by overlapping sub-transaction execution. Third, there is a substantial gap between \textsf{partially-async} and \textsf{fully-async}, due to asymmetric costs between receiving procedure results and sending procedure invocations to other reactors. The latter manifests because of thread switching costs across cores in the receive code path, as opposed to atomic operations in the send code path.
In \textsf{opt}, latency is further reduced when compared to \textsf{fully-async} by cutting the processing costs almost in half, which have a smaller impact than communication across cores. It is interesting to note that these optimizations can be done on the $\mu$sec scale. The programming model allows a developer to reduce the latency of a transaction from 86~$\mu$sec to 25~$\mu$sec by simple program reformulations without compromising~consistency.

\subsubsection{Cost Model Breakdown}
\label{sec:exp:pred:programs:breakdown}
In this section, we break down our measurements of transaction latencies by the cost components in Figure~\ref{fig:cost:model}, and further validate that our cost model is realizable in \reactdb. We focus on the \textsf{fully-sync} and \textsf{opt} multi-transfer formulations described above, and vary the size of the multi-transfer transaction by changing the number of destination accounts similarly to Figure~\ref{fig:latency:programs}. For each variant, we profiled the execution time of the programs in \reactdb\ into the components of the cost model of Figure~\ref{fig:cost:model}. In addition, we used the profiling information from \textsf{fully-sync} for a transaction size of one to calibrate the parameters of the cost model for prediction, including processing and communication costs. From the parameter values for the single-transfer \textsf{fully-sync} run, we employed the cost equation of Figure~\ref{fig:cost:model} to predict the execution costs for other transaction sizes and for both the \textsf{fully-sync} and \textsf{opt} program formulations. The predicted values are labeled \textsf{fully-sync-pred} and \textsf{opt-pred}.

We break down transaction latencies into the following components: (a)~\emph{sync-execution}: the cost of processing the logic in the transaction and in synchronous sub-transactions, corresponding to the first two components of the cost equation in Figure~\ref{fig:cost:model}; (b)~$C_s$ and $C_r$: the forward and backward costs of communication between reactors in the third component of the cost equation; (c)~\emph{async-execution}: the cumulative execution cost of all asynchronous sub-transactions overlapped with any synchronous sub-transactions and processing logic, corresponding to the fourth component of the cost equation; (d)~\emph{commit + input-gen}: the cost of the commit protocol, including OCC and 2PC, along with the time to generate the inputs for the transaction. The latter cost component is not shown in Figure~\ref{fig:cost:model} since it only applies to root transactions and not to any sub-transaction in the reactor programming model. As such, we would expect the bulk of the difference between the predicted and observed performance to be explainable by this cost component. 

Figure \ref{fig:latency:cost:model} shows that the predicted breakdown for the cost components closely matches the latencies profiled for actual executions in \reactdb, even at such a fine granularity. The slight difference in the overlap of different bars is within the variance of the observed vs. the calibration measurement, and expected especially since calibration measures parameters within the 5$\mu$sec range. For a transaction size of one, we can see that \textsf{opt} has the same performance behavior as \textsf{fully-sync}. This effect arises because the destination transaction executor for the credit in the transfer is the same as the source transaction executor for the debit, resulting in a synchronous execution of the credit and debit sub-transactions similar to \textsf{fully-sync}. As we increase the transaction size, the number of transaction executors spanned by the transaction increases, and the execution costs of \textsf{fully-sync} grow because of increasing costs in \emph{sync-execution}, $C_s$ and $C_r$. We again observe here cost asymmetry between $C_s$ and $C_r$, arising for the same reasons remarked in Section~\ref{sec:exp:pred:programs}. 
For \textsf{opt}, we do not observe any \emph{sync-execution} costs, since all credit sub-transactions are overlapped with each other and with the single debit on the source reactor. The growth in the \emph{async-execution} cost of \textsf{opt} with increasing transaction size is caused by the rising communication cost for the sequence of credit sub-transactions, i.e., the last asynchronous credit sub-transaction incurs a cumulative cost of communication of all asynchronous sub-transactions before it in the sequence. 

In summary, we observe that  
the latencies of transactions can be reliably profiled in our system \reactdb\ into the components of the cost model of Figure~\ref{fig:cost:model} even in a challenging scenario where the cost of the processing logic in the benchmark is extremely small and comparable to the cost of communication across reactors. This evidence indicates that it is possible for developers to observe and explain the latency behavior of programs with reactors and to reformulate their programs for better performance. Further evaluation of the effect of physical configuration on latency is provided in Appendix~\ref{sec:exp:pred:cost}.

\begin{figure*}[!t]
  \begin{minipage}{0.32\linewidth}
    \centerline{\includegraphics[width=0.95\linewidth]{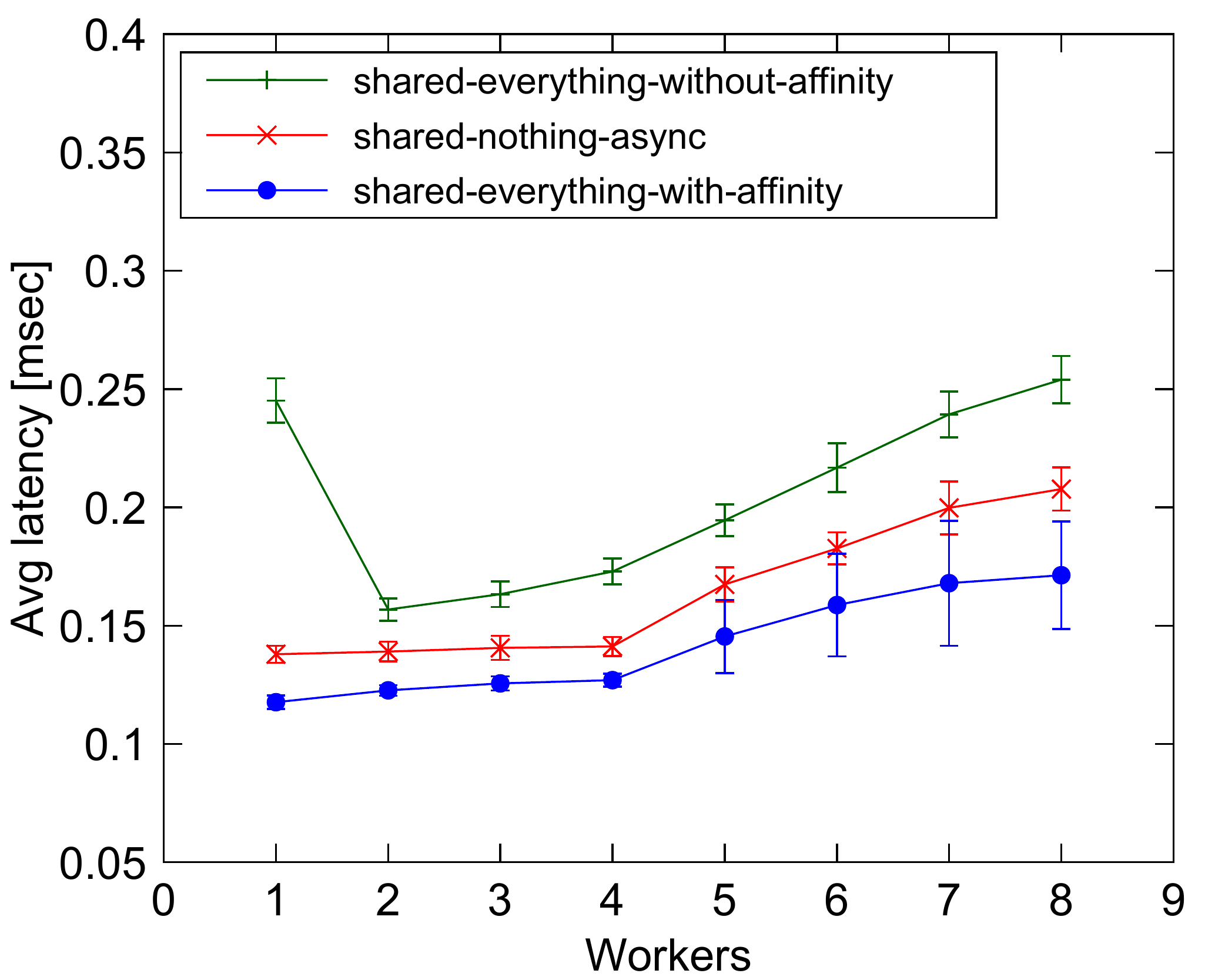}}
    \vspace{-1ex}
    \caption{TPC-C latency with varying load at scale factor 4.}\label{fig:latency:fixed-db-load:deployments}
  \end{minipage} \hfill
  \begin{minipage}{0.32\linewidth}
    \centerline{\includegraphics[width=0.95\linewidth]{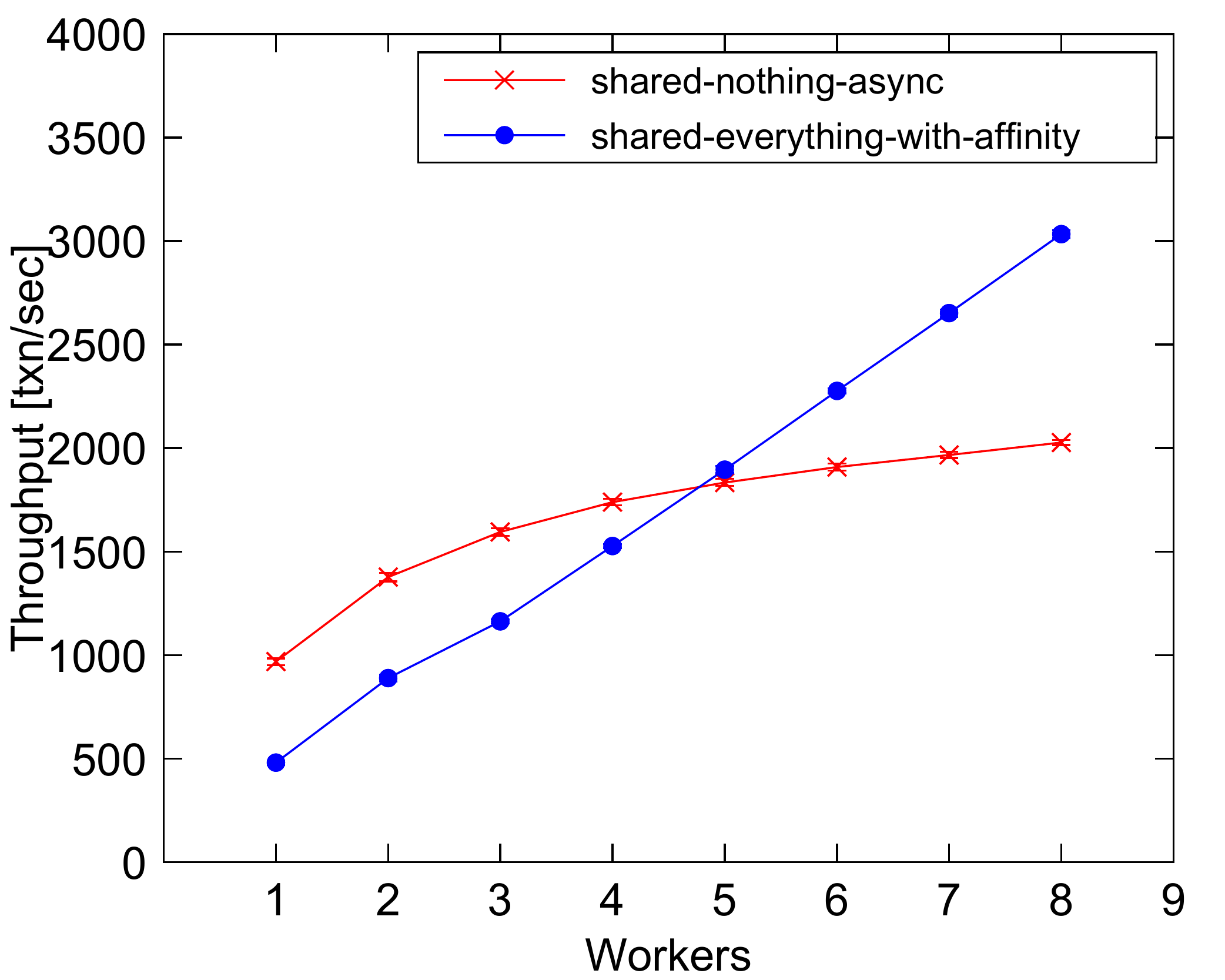}}
    \vspace{-1ex}
    \caption{Throughput of new-order-delay transactions with varying load.} \label{fig:throughput:load:deployments}
  \end{minipage} \hfill
  \begin{minipage}{0.32\linewidth}
    \centerline{\includegraphics[width=0.95\linewidth]{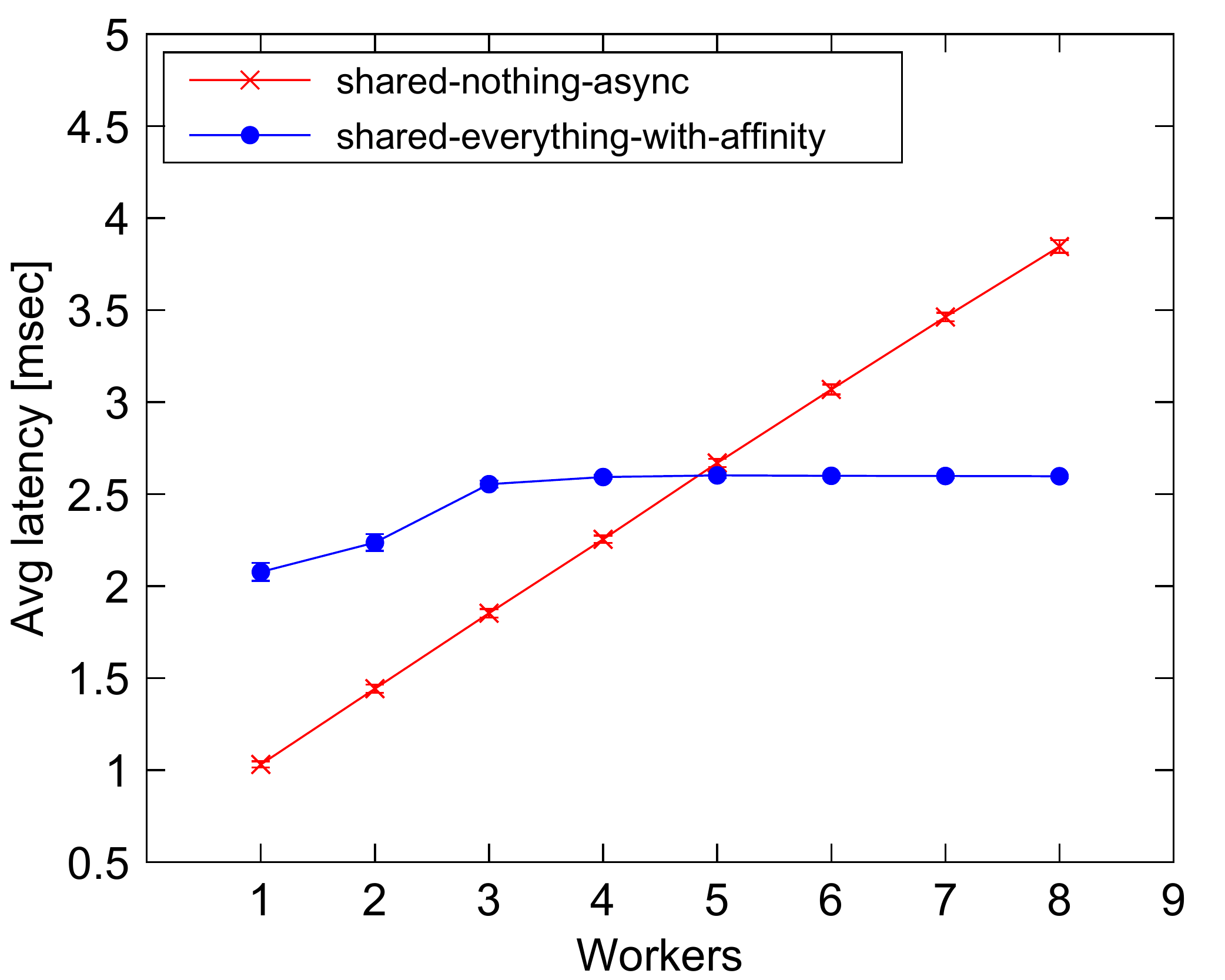}}
    \vspace{-1ex}
    \caption{Latency of new-order-delay transactions with varying load.} \label{fig:latency:load:deployments}
  \end{minipage} \hfill
\end{figure*}

\subsection{Virtualization of Database Architecture}
\label{sec:exp:virt}
\subsubsection{Effect of Load}
\label{sec:load:fixed:db}
We evaluate the behavior of the three database architecture deployments described in Section~\ref{sec:deployments} with the standard mix of the TPC-C benchmark under increasing load from client workers. We fixed the database size to a scale factor of four, corresponding to four warehouse reactors. Correspondingly, we deployed four transaction executors in all configurations, but increased the number of workers invoking transactions on the database from 1 to 8. As such, we expect the database to get progressively more overloaded from 5 to 8 workers.

Figures~\ref{fig:throughput:fixed-db-load:deployments} and~\ref{fig:latency:fixed-db-load:deployments} show the average throughput and latency of successful transactions in this scenario. We observe that \textsf{shared-everything-with-affinity} outperforms the other deployment configurations, since it exploits memory access affinities and minimizes cross-core communication overheads. Even though \textsf{shared-nothing-async} attempts to overlap cross-reactor sub-transactions as much as possible, the relative cost of dispatching sub-transactions with limited processing logic makes it more expensive than \textsf{shared-everything-with-affinity}, which employs direct memory accesses. We observe that the differences in throughput between the two methods are not large between 1 and 4 workers, since there is only a 1\% chance of items for stock update in the new-order transaction and a 15\% chance of customer lookups in the payment transaction being remote. Since workers employ client affinity to warehouses, note that for 5 workers, there are two worker threads generating workload for the first warehouse; for eight workers, there are two worker threads generating workload for every warehouse. So from 5 to 8 workers, the possibility of conflicts between concurrent transactions to the same warehouses arises, especially for the payment and new-order transactions.

As expected, therefore, from 1 to 4 workers, abort rates for all deployments are negligible and close to nil. Abort rates then go up to 4.1\% for \textsf{shared-everything-without-affinity} and 5.72\% for \textsf{shared-nothing-async} for eight workers. Interestingly, \textsf{shared-everything-with-affinity} is resilient to this effect, because each transaction executor runs a transaction to completion before picking up the next one from its queue. In other words, since \textsf{shared-everything-with-affinity} does not employ asynchronicity, transaction executor threads never block, and executors can operate with a multi-programming level of one. At the same time, affinity in routing ensures that each transaction executor services transactions for a different warehouse reactor, preserving identical execution behavior to that observed for workers 1 to 4 despite changes in the conflict patterns in workload generation. Even though \textsf{shared-nothing-async} also ensures affinity in transaction routing, it uses cooperative multi-threading and a higher multi-programming level for greater hardware utilization, and ends up executing multiple transactions concurrently from a given transaction executor's queue.  

While we increase the number of client workers, we must remember that the database system processing resources in the form of transaction executors remain fixed. So the throughput on average increases for all the deployments because the hardware utilization goes up. The hardware utilization on each of the transaction executor cores for \textsf{shared-everything-with-affinity} increases from 0 to 83\% between 1 and 4 workers, and is pushed further to 99\% as we get to eight workers. Given the use of asynchronicity in \textsf{shared-nothing-async}, this deployment uses all cores from the start, with a hardware utilization for one worker on the four transaction executor cores of 76\%, 2.5\%, 2.5\% and 2.5\%, respectively. The hardware utilization of transaction executor cores with this deployment is uniform at 79\% with four workers, and keeps on rising up to 98\% for eight workers. For \textsf{shared-everything-without-affinity}, the hardware utilization is uniform throughout, rising from 17\% to 66\% from 1 to 4 workers and reaching only 84\% with eight workers.

We observe that \textsf{shared-everything-without-affinity} exhibits the worst performance. At one worker, every transaction invocation is routed by the round-robin load balancing router to a different transaction executor than the one that processed the last request, amplifying cross-core communication. As workers are added this effect diminishes; however, the lower hardware utilization and the eventual rise in abort rates limits the efficiency of this architecture. 

In short, we remark that the capability to virtualize database architecture allows \reactdb\ to be configured to maximize hardware utilization and minimize conflicts in a standard OLTP benchmark. Moreover, we observe that asynchronicity in transactions engenders a trade-off between communication overheads and processing costs. We validate this observation by fitting our cost model to the TPC-C new-order transaction in Appendix~\ref{sec:cost:new:order}, and explore asynchronicity trade-offs further in the next section.

\subsubsection{Asynchronicity Tradeoffs}
\label{sec:load}
To additionally drill down on the potential benefits of asynchronicity under concurrency, we evaluate in this section the two database architectures \textsf{shared-nothing-async} and \textsf{shared-everything-with-affinity} under varying load. We control the amount of load by varying the number of workers from 1 to 8, while keeping the number of warehouses constant at a scale factor of eight. 
For clarity, we focus exclusively on new-order transactions. Each new-order consists of between 5-15 items and we force each of the items to be drawn from a remote warehouse with equal probability. Since the default new-order formulation has limited parallelism in the logic executed at remote warehouses, we augmented the logic for stock data update with an artificial delay between 300 and 400~$\mu$sec by generating random numbers to model stock replenishment calculations. This increases the overall work in the transaction without increasing its data footprint and the contention on the database. 

\begin{sloppypar}
Figures~\ref{fig:throughput:load:deployments} and~\ref{fig:latency:load:deployments} show the throughput and latency, respectively, of running 100\% new-order-delay transactions under increasing load. With one worker, the throughput of \textsf{shared-nothing-async} is double that of \textsf{shared-everything-with-affinity}. The former executes all the stock updates across ~5-6 remote warehouse asynchronously (average distinct remote warehouses chosen from 7 using a uniform distribution) fully utilizing the available hardware parallelism, while the latter executes the entire transaction logic sequentially. Although \textsf{shared-nothing-async} incurs higher communication cost in dispatching the stock updates to be performed by different warehouse reactors, the greater amount of work in each stock update makes it worthwhile in comparison to sequential shared memory accesses in \textsf{shared-everything-with-affinity}. Conversely, as we increase the number of workers and thus pressure on resources, the throughput of \textsf{shared-nothing-async} starts growing less than that of \textsf{shared-everything-with-affinity}. 
Note that the abort rate for the deployments was negligible (0.03-0.07\%), highlighting the limited amount of contention on actual items. 
\end{sloppypar}

In summary, these results suggest that the most effective database architecture may change depending on load conditions when asynchronicity can be exploited by transaction code. Under high load, \textsf{shared-everything-with-affinity} exhibits the best performance among the architectures evaluated, since it reduces overhead at the expense of not utilizing at all intra-transaction parallelism. On the other hand, when load conditions are light to normal and when transaction logic comprises enough parallelism, \textsf{shared-nothing-async} can achieve substantially higher throughput and lower latency. To further validate these observations, we evaluate in Appendix~\ref{sec:cross:reactor} the effects of varying cross-reactor accesses in the TPC-C benchmark under conditions of high load.

\section{Related Work}
\label{sec:related}

\subsection{In-memory OLTP Databases}
H-Store~\cite{Stonebraker:2007:EAE:1325851.1325981} and HyPer~\cite{Kemper:2011:HHO:2004686.2005619} follow an extreme shared-nothing design by having single-threaded execution engines responsible for each data partition. As a result, single-partition transactions are extremely fast, but multi-partition transactions and skew greatly affect system throughput. LADS~\cite{YaoA0LOWZ16:Lads} improves upon this limitation by merging transaction logic and eliminating multi-partition synchronization through dynamic analysis of batches of specific transaction classes. In contrast to these shared-nothing engines, shared-everything lock-based OLTP systems specifically designed for multi-cores, such as DORA~\cite{Pandis:2010:DTE:1920841.1920959} and PLP~\cite{Pandis:2011:PPL:2021017.2021019}, advocate partitioning of internal engine data structures for scalability.  Orthrus~\cite{RenFA16:Orthrus} partitions only the lock manager and utilizes a message-passing design for lock acquisition to reduce lock contention across multiple cores for contended workloads.

In contrast to the baked-in architectural approach of earlier engines, \reactdb\ borrows the highly-scalable OCC implementation of Silo~\cite{Tu:2013:STM:2517349.2522713}, building on top of it a virtualization layer that allows for flexible architectural deployments, e.g., as a classic shared-everything engine, a shared-nothing engine, or an affinity-based shared-everything engine. In addition, \reactdb\ is not restricted to specific transaction classes, supporting transactions with, e.g., user-defined aborts, conditionals, and range queries. Finally, \reactdb\ is the first engine realizing the programming model of reactors.

\subsection{Transactional Partitioned Data Stores}
A class of systems provides transactional support over key-value stores as long as keys are co-located in the same machine or key group~\cite{DasAA10:GStore, DasAA09:Elastras}. Warp~\cite{EscrivaWS15:Warp}, in contrast, provides full transaction support with nested transactions, but limits query capabilities, e.g., no predicate reads are provided nor relational query support. The limited transactional support and low-level storage-based programming model make it difficult to express applications as opposed to the reactor programming model, which provides serializable transactions with relational query capabilities.~Recent work has also focused on enhancing concurrency through static analysis of transaction programs~\cite{MuCZLL14:Rococo,ZhangPZSAL13:Lynx}. 
The latter could be assimilated in the implementation of \reactdb's concurrency control layers as future work.

\subsection{Asynchronous Programming}
As mentioned previously, reactors are a novel restructuring in the context of databases of the actor model~\cite{Agha:1986:Actors}.
In contrast to regular actors, reactors comprise an explicit memory model with transactions and relational querying, substantially simplifying program logic. 
These features make the reactor model differ significantly from the virtual actors of Orleans~\cite{BernsteinBGKT14:Orleans} and from other actor-based frameworks~\cite{Armstrong10:Erlang,HaydukSF15:STMActor}. Recent work in Orleans has focused on a vision of integrating traditional data-management functionality in a virtual actor runtime for the middle tier of a classic three-tier architecture~\cite{BernsteinDKM17:ActorDB,transactions-distributed-actors-cloud-2}. This approach is complementary to our work of integrating actor features in a database system, i.e., enriching the data tier itself. Additionally, \reactdb\ comprises building a high-performance, scalable, multi-core OLTP system with an actor-oriented programming model and latency control, which is not the target design and feature set of the vision for Orleans~\cite{BernsteinDKM17:ActorDB}.

As explained in Section~\ref{sec:programming:model}, reactors are related to the early work on Argus~\cite{LiskovCJS87:Argus} because of the asynchronous transactional programming model supporting nested function calls; however, the reactor programming model is substantially different from that of Argus. First, the use of a relational data and query model is a central idea of reactors, but not of Argus. Note that the latter is not a simple restriction of the former, because the programming issues handled by a relational abstraction, e.g., physical data independence, 
would need to be coded from scratch at a very low level in Argus. Second, user-defined logical actors are a central idea of reactors, but not of Argus. Guardians in Argus employ multiple concurrent processes explicitly, while reactors abstract away the mapping to threads in \reactdb. 
Third, reasoning about latency from the programming model is a central idea of reactors, but again not of Argus. Even though Argus has low-level asynchronous calls, it lacks an explicit cost model of synchronous and asynchronous communication. On the system implementation level, \reactdb\ is an OLTP database system designed for low-overhead virtualization of database architecture, which was never the focus of Argus. These differences to Argus also distinguish our work from a large class of object-oriented distributed computing and operating systems~\cite{LazowskaLAFFV81:Eden, Birman85:Isis,DasguptaLA88:Cloud, ChrysanthisRSV86:Gutenberg, Oakley89:Mercury}.

\subsection{Database Virtualization}
Virtualization of database engines for cloud computing has focused on particular target database architectures, e.g., shared-nothing databases with transactional support only within partitions~\cite{Bernstein:2011:SQLAzure} or distributed control architectures with weaker consistency guarantees~\cite{KossmannKL10:EvaluationCloudDB}. By contrast, \reactdb\ offers infrastructure engineers the possibility to configure database architecture itself by containerization, while maintaining a high degree of transaction isolation. Our results support recent observations of low overhead of use of container mechanisms together with an in-memory database~\cite{MuhlbauerRKK015:LowOverheadContainer}, while showing that even more flexibility in database architecture can be achieved at negligible cost.  

The design of \reactdb\ is reminiscent of work on OLTP on hardware islands~\cite{PorobicPBTA12:HardwareIslands} and on the cost of synchronization primitives~\cite{David:2013:Synchronization}. 
Complementary to recent work on compiler optimizations~\cite{Ramachandra:2017:Froid}, reactors can be seen as a step towards improving programmability and support for stored procedures in database systems.

\section{Conclusion}
In this paper, we introduced reactors, a new relational abstraction for in-memory OLTP databases. Reactors comprise an asynchronous programming model that allows encoding of intra-transaction parallelism in database stored procedures while maintaining serializability. We presented the design of \reactdb, the first implementation of reactors. \reactdb\ 
enables flexible and controllable database architecture configuration at deployment time. 
Reactors open up a variety of directions for future work, ranging from reactor database modeling to efficient mapping of reactors to distributed hardware architectures.

\begin{acks}
We thank the anonymous reviewers for their insightful comments and Phil Bernstein for discussions and suggestions, which have helped us improve this paper.
\end{acks}

\bibliographystyle{ACM-Reference-Format}
\bibliography{reactdb}
\appendix
\begin{figure*}[!t]
  \begin{minipage}{0.32\linewidth}
    \centerline{\includegraphics[width=0.8\linewidth]{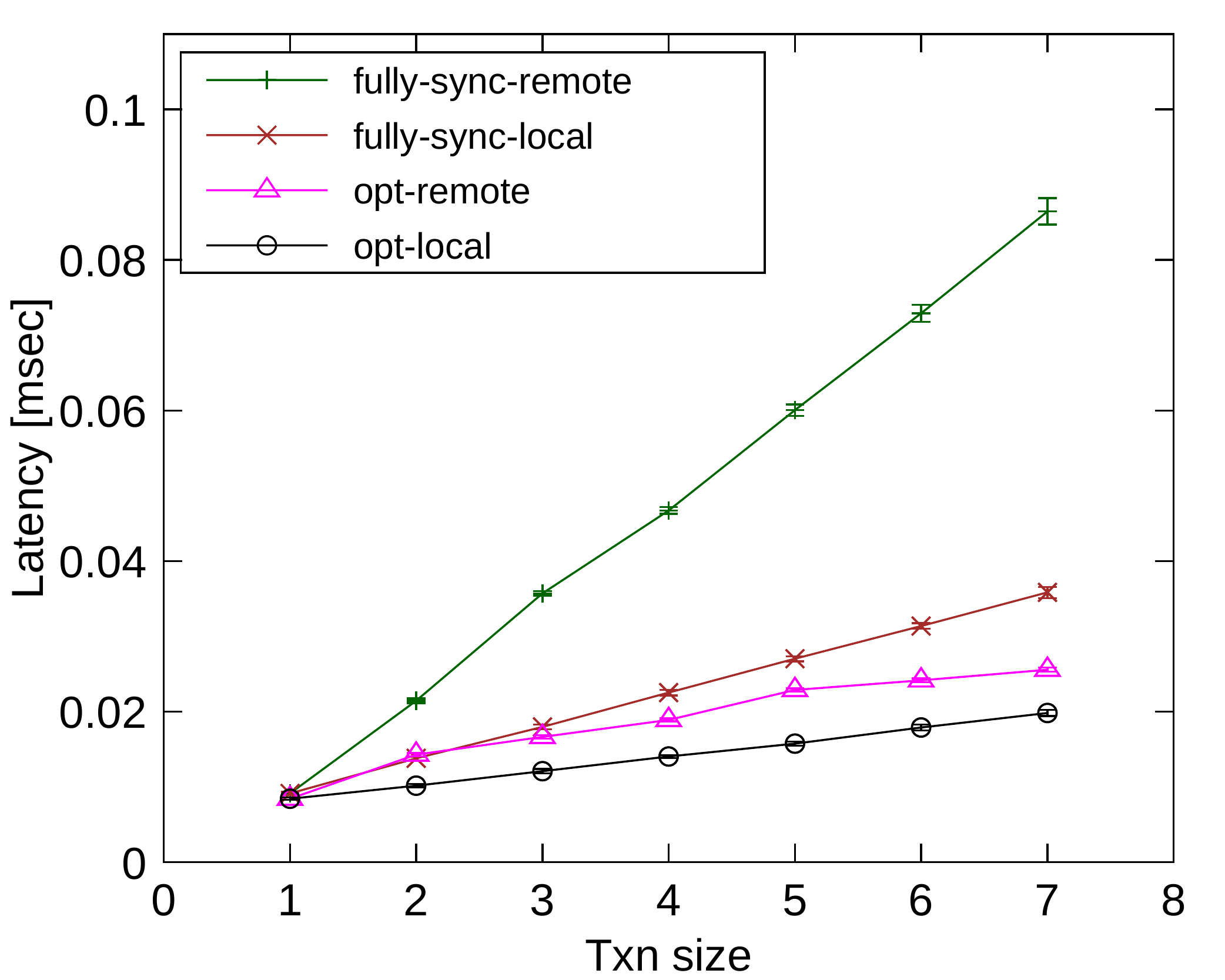}}
        \caption{Latency vs. size and different target reactors spanned.}\label{fig:latency:txnsize}
  \end{minipage} \hfill
  \begin{minipage}{0.32\linewidth}
    \centerline{\includegraphics[width=0.8\linewidth]{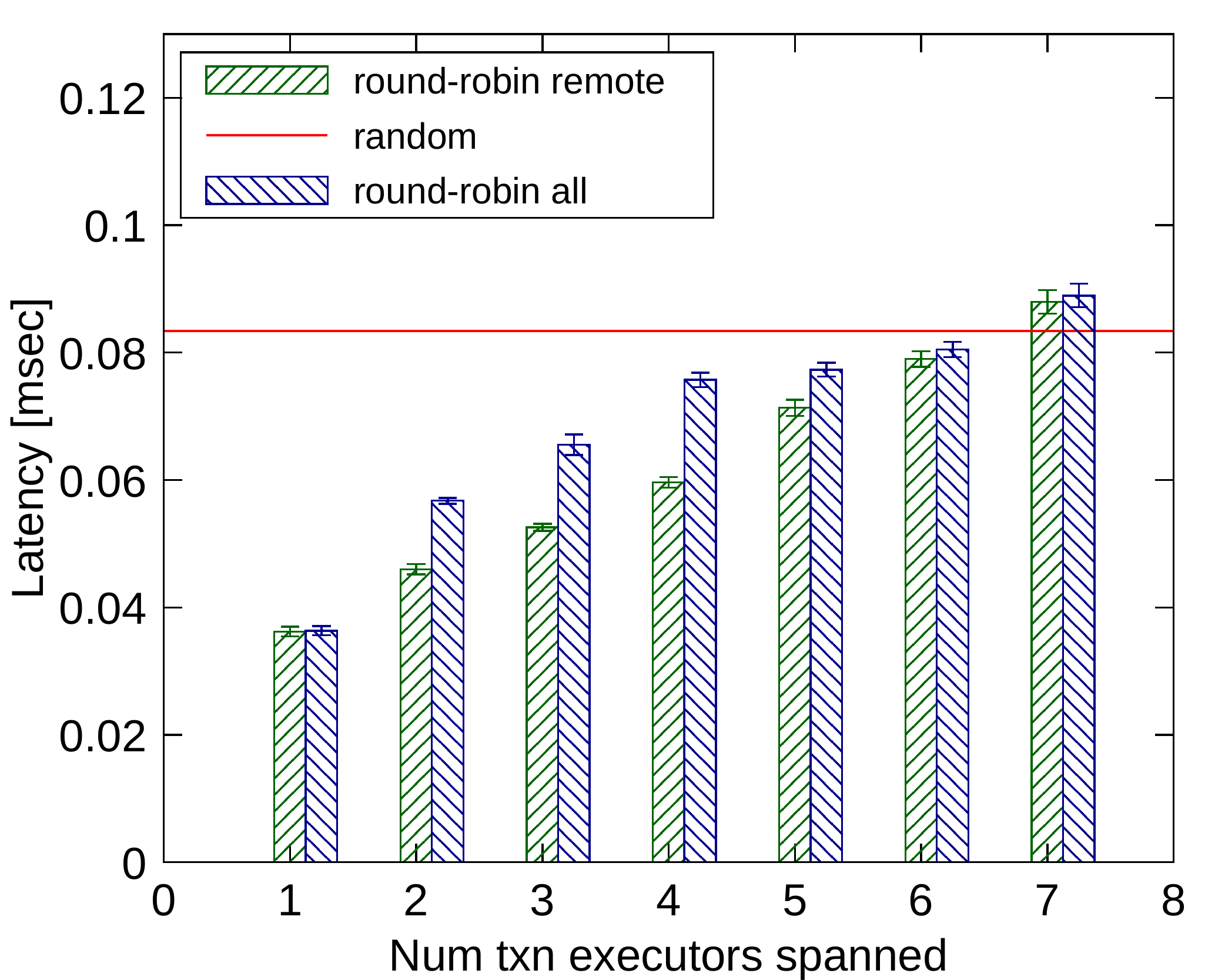}}  
    \caption{Latency vs. distribution of target reactors and fixed size.}\label{fig:latency:dist:degree}
  \end{minipage} \hfill
  \begin{minipage}{0.32\linewidth}
    \centerline{\includegraphics[width=0.8\linewidth]{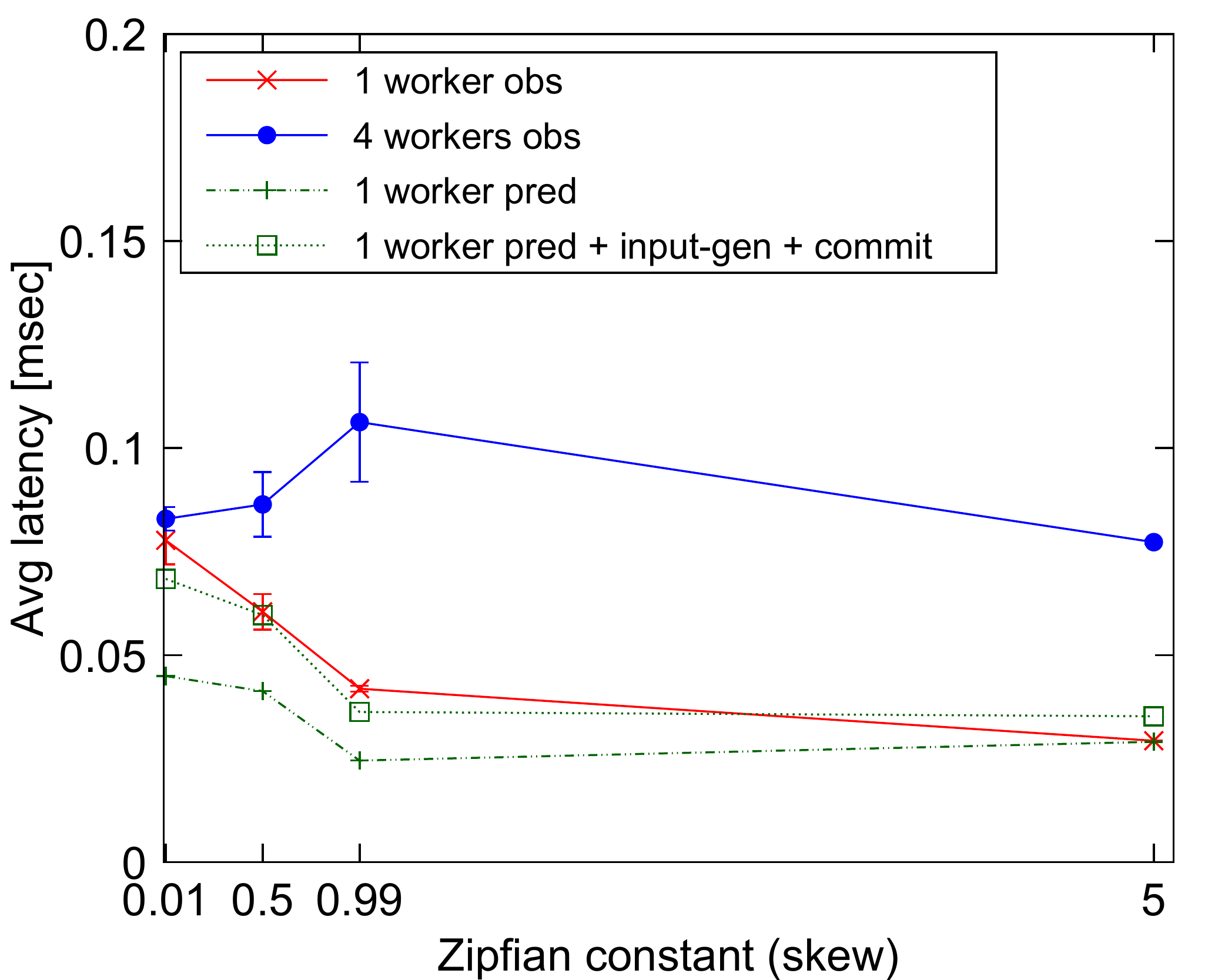}}
    \caption{Effect of skew and queuing on latency.} \label{fig:cost-model:skew}
  \end{minipage} \hfill
\end{figure*}

\section{Proof of Theorem~\ref{THM:EQUIVALENCE}}
\label{thm:equivalence:proof}

Let us assume $H$ is serializable and $H'$ is not serializable. From the serializability theorem, since H is serializable, the serializability graph of $H$ ($SG(H)$) is acyclic; since the projected history $H'$ is not serializable, the serializability graph $SG(H')$ must be cyclic. Therefore, there must exist a cycle $T'_{i} \rightarrow \ldots \rightarrow T'_{j} \rightarrow \ldots \rightarrow T'_{i}$. Since the graph is built on operations of the classic transactional model, then there must be conflicting operations $o'_{i} <_{P_H} \ldots <_{P_H} o'_{j} <_{P_H} \ldots <_{P_H} o'_{i}$. By condition (3) of Definition~\ref{history-tim}, there must exist sub-transactions $ST_{i,l}^{k} \in T_i$ and $ST_{j,l'}^{k'} \in T_j$ such that $ST_{i,l}^{k} <_{H} \ldots <_{H} ST_{j,l'}^{k'} <_{H} \ldots <_{H} ST_{i,l}^{k}$. As a result, $SG(H)$ must be cyclic, and we arrive at a contradiction. To show the reverse direction, it is simple to follow a similar argument, but starting with a cyclic graph in $SG(H)$ and showing that $SG(H')$ must be cyclic as well in contradiction. 

\section{Latency Control across Physical Configurations}
\label{sec:exp:pred:cost}

In this appendix, we complement the results in Section~\ref{sec:exp:pred} by evaluating the latency impact of different physical mappings of reactors in \reactdb. In particular, we provide additional experiments showing that observed latencies for the multi-transfer transaction can be reliably explained by whether the affected reactors are co-located or not on physical components employed in a given configuration.
\subsection{Local vs. Remote Calls}
In this section, we show how the configuration of physical distribution can affect the latency of transactions. The cost equation of Figure~\ref{fig:cost:model} models communication costs among reactors (namely $C_s$ and $C_r$), which are higher when reactors are mapped to containers over distinct physical processing elements. We term calls among such physically distributed reactors \emph{remote calls}. 
By contrast, calls between reactors mapped to the same container are termed \emph{local~calls}.

To highlight cost differences in remote calls, we consider the \textsf{fully-sync} and \textsf{opt} multi-transfer formulations.~We evaluate two extremes: either destination accounts span all containers (\textsf{-remote}) or are on the same container as the source account~(\textsf{-local}).~Figure \ref{fig:latency:txnsize} shows that the cost of \textsf{fully-sync-remote} rises sharply because of increase in both processing and communication costs compared to \textsf{fully-sync-local}, which only sees an increase in processing cost.~There is a comparatively small difference between \textsf{opt-local} and \textsf{opt-remote}, since the processing of remote credit sub-transactions is overlapped with the local debit sub-transaction, and thus part of the fourth component summed in Figure~\ref{fig:cost:model}. The extra overhead in \textsf{opt-remote} comes from larger, even if partially overlapped, communication and synchronization overheads to invoke the sub-transactions on the remote transaction executors and receive results. 

\subsection{Varying Degree of Physical Distribution}
In order to better understand the growth in communication costs due to remote calls, we conduct another experiment where we fix the multi-transfer transaction size to seven destination accounts, and then control these accounts so as to span a variable number of containers. Recall that in the deployment for this experiment, each of the seven containers has exactly one transaction executor pinned to a hardware thread. We use the \textsf{fully-sync} formulation of multi-transfer, so we expect to see higher latencies as a larger number of the credits to the seven destination accounts are handled by remote transaction executors.  

We experiment with three variations for selecting destination accounts for our multi-transfer transaction as we vary the number $k$ of transaction executors spanned from one to seven. The first variant, \textsf{round-robin remote}, performs $7-k+1$ local debit calls by choosing accounts mapped to the first container, and $k-1$ remote calls by choosing accounts round-robin among the remaining containers. The second variant, \textsf{round-robin all}, performs $\lceil 7/k \rceil$ local calls and $\lfloor 7/k \rfloor$ remote calls. Finally, we measure an expected value for latency by selecting destination accounts with a uniform distribution, termed \textsf{random}.   
 
Figure~\ref{fig:latency:dist:degree} shows the resulting latencies.~We observe a smooth growth in the latency for \textsf{round-robin remote}, since we increase the number of remote calls exactly by one as we increase the number of transaction executors spanned. The behavior for \textsf{round-robin all} differs in an interesting way. For two transaction executors spanned, \textsf{round-robin all} performs three remote calls and four local calls. While \textsf{round-robin all} performs four remote calls and three local calls for three transaction executors, the method performs five remote calls and two local calls for four to six transaction executors spanned. These effects are clearly reflected in the measured latencies. For \textsf{random}, the expected number of remote calls is between six and seven, which is again tightly confirmed by the latency of the multi-transfer transaction in Figure \ref{fig:latency:dist:degree}.

\section{Effect of Skew and Queuing on Latency}
\begin{figure*}[!t]
   \begin{minipage}{0.32\linewidth}
    \centerline{\includegraphics[width=0.8\linewidth]{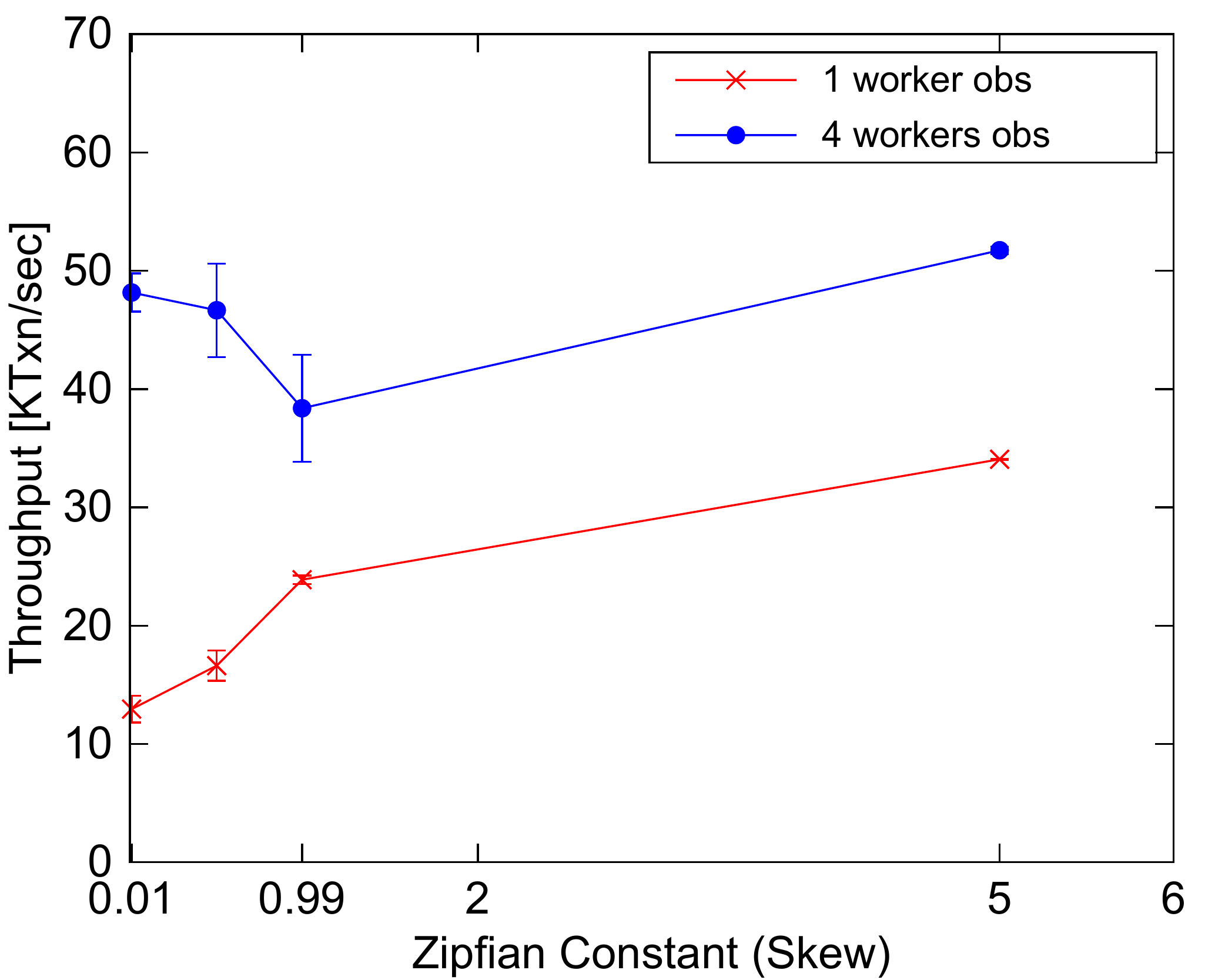}}
    \caption{Effect of skew and queuing on throughput.} \label{fig:cost-model:skew:throughput}
  \end{minipage} \hfill
  \begin{minipage}{0.32\linewidth}
    \centerline{\includegraphics[width=0.8\linewidth]{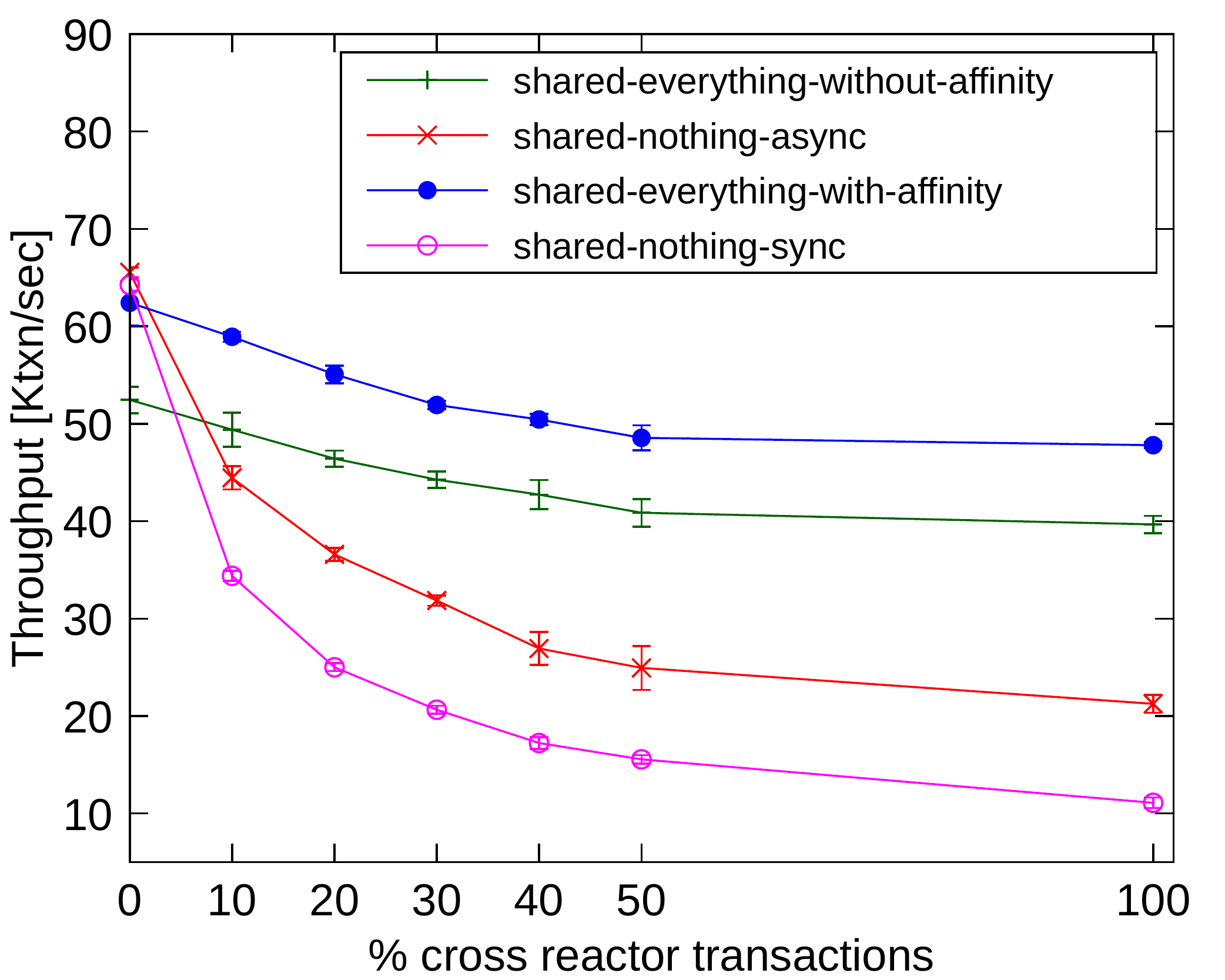}}
    \caption{Throughput of cross-reactor TPC-C new-order (scale factor~8).} \label{fig:throughput:dist:degree:deployments}
  \end{minipage} \hfill
  \begin{minipage}{0.32\linewidth}
    \centerline{\includegraphics[width=0.8\linewidth]{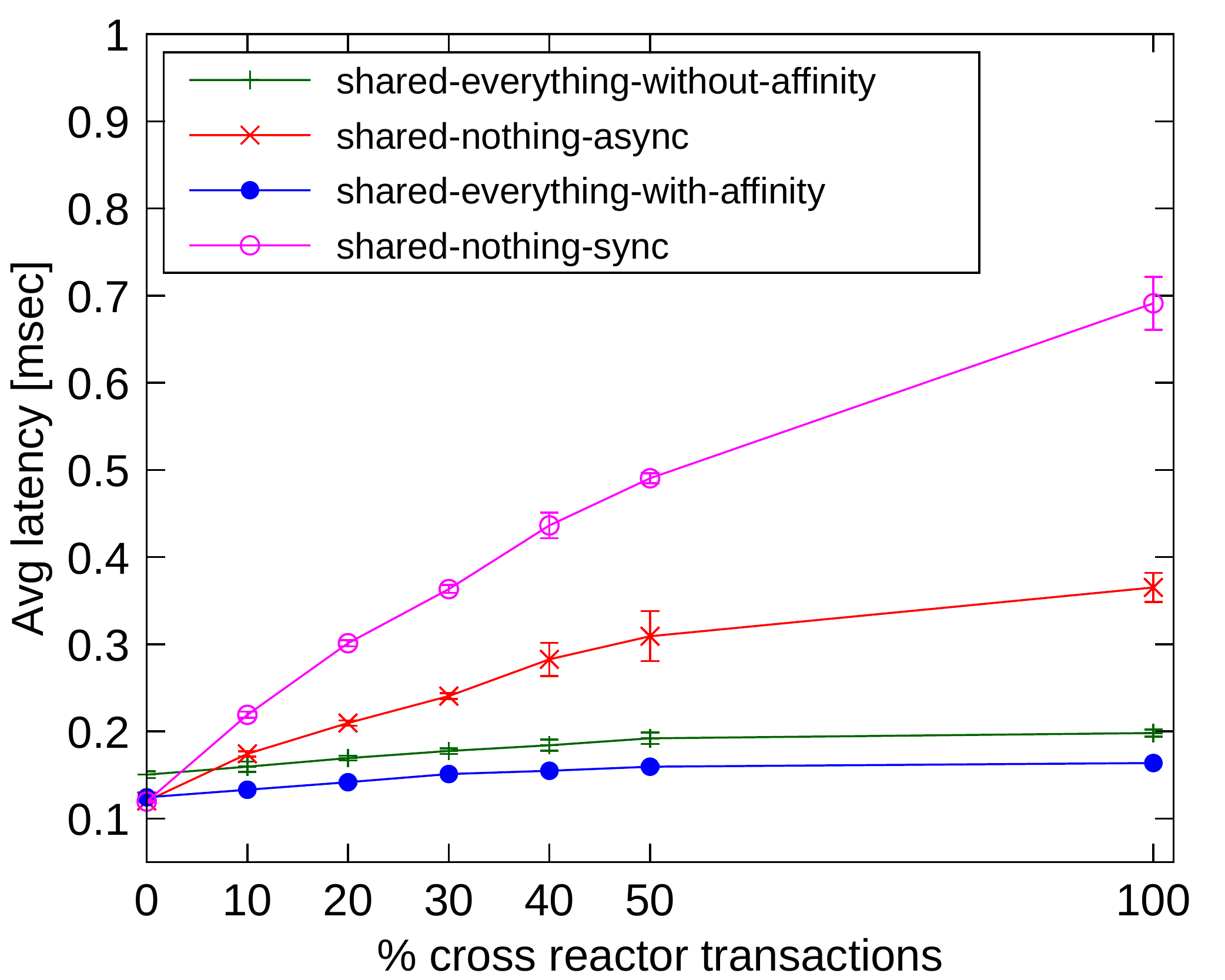}}
    \caption{Latency of cross-reactor \\ TPC-C new-order (scale factor~8).} \label{fig:latency:dist:degree:deployments}
  \end{minipage} \hfill
\end{figure*}
\label{sec:exp:pred:limitations}
In this appendix, we drill down experimentally into the limitations discussed for our cost model in Section~\ref{sec:perf:predict}, evaluating a scenario with queuing delays and skew. 

\minisec{Setup}
We created a separate setup to evaluate limitations in explanatory ability of our cost model by augmenting the YCSB benchmark~\cite{CooperSTRS10:YCSB} with a \texttt{multi\_update} transaction that updates 10 keys. 
We model each key as a reactor, and the \texttt{multi\_update} transaction invokes a read-modify-write update sub-transaction asynchronously for each key (record size of 100 bytes). Unlike in the setup with Smallbank, we configure \reactdb\ such that transactions may not realize full parallelism at the system level. In particular, we chose a scale factor of 4 where each scale factor corresponds to 10,000 keys. Four containers are deployed, each with one transaction executor, and assigned 10,000 contiguous reactors corresponding to the scale factor. Furthermore, we select the reactor where the \texttt{multi\_update} transaction is invoked randomly from the set of 10 keys in the update, and choose the keys for \texttt{multi\_update} from a zipfian distribution. As such, we model program logic where it is not possible to determine statically how many sub-transactions will be realized by the system synchronously or asynchronously. However, to ensure that transactions remain fork-join, we sorted the keys to be updated in a \texttt{multi\_update} transaction such that keys deployed in remote transaction executors precede the ones in the local transaction executor where the transaction is initiated. Finally, the experiment is performed both with one and four workers to simulate absence and presence of queuing delays, respectively. 

\minisec{Results}
We employ the \texttt{multi\_update} transaction under a 100\% mix.  Recall that our setup explained above explicitly restricts the parallelism that can be realized by the system. So to be able to fit our cost model, we recorded the average sizes of the realized $\text{sync}_{ovp}(ST_{i,j}^{k})$ and $\text{async}(ST_{i,j}^{k})$ sequences with the zipfian distribution employed. Similar to the experiment in Section~\ref{sec:exp:pred:programs:breakdown}, we calibrated the average communication and processing cost parameters by profiling the \texttt{multi\_update} transaction with updates to a single key chosen from a uniform distribution. We emphasize that the cost model abstracts intra-transaction parallelism to aid developers in contrasting transaction program formulations, and thus does not capture interference among concurrent workers or queueing effects. As such, we expect cost model predictions to be most accurate in configurations where a single worker is deployed, as in Section~\ref{sec:exp:pred}. 

Figures~\ref{fig:cost-model:skew} and~\ref{fig:cost-model:skew:throughput} show the observed behavior of average transaction latency and throughput, respectively, with varying skew as captured by the zipfian constant used to select keys for \texttt{multi\_update}, for different number of workers. Figure~\ref{fig:cost-model:skew} has been augmented by the cost model predictions of transaction latencies. For one worker, the observed average latency decreases as skew increases, given that more of the sub-transactions invoked in \texttt{multi\_update} become synchronous. This effect arises since the communication overhead to dispatch a sub-transaction to a remote reactor is greater than the time to process a single update. This trend is also captured in our cost model prediction, \textsf{1 worker pred}, which shows decreasing average latency as the zipfian constant increases to 0.99. 
Interestingly, the curve predicts that the average latency should increase after 0.99, which conflicts with our observation. This is because the cost of processing actually decreases when all updates are to the same key in the transaction. In this case, the read in a read-modify-write can just return the previous updated value from the transaction's write-set instead of accessing the index structure, which was not accounted for while calibrating processing cost. 
In addition to \textsf{1 worker pred}, we also show a curve adding to the prediction the measured costs of input generation and commitment. We remind the reader that the latter two costs are not part of the equation in Figure~\ref{fig:cost:model}, and as in Section~\ref{sec:exp:pred:programs:breakdown}, the difference between predicted and observed latencies for a single worker are largely accounted by these two factors. 

With four workers, queueing and skew together lead to both higher and more variable latencies overall, which are as expected not captured by the cost model. 
Abort rates also increase, going from 0.26\% to 3.24\% as the zipfian constant increases from 0.01 to 0.99. Hardware utilization on the transaction executor core hosting the most accessed reactors reflects the trend in the figure, corresponding to 63\%, 92\% and 100\% at zipfian constants 0.01, 0.5 and 0.99. At 5.0 skew, a single reactor is accessed, eliminating variability but not queueing. 

\section{Cost Model Validation with TPC-C New-Order}
\label{sec:cost:new:order}

\begin{table}
  \small
\begin{tabular}{|c|c|c|c|c|c|c|}
  \hline
   \multirow{4}{1cm}{\centering \textbf{Cross Reactor Access \%}}& \multicolumn{4}{c|}{\textbf{1 worker}} & \multicolumn{2}{c|}{\textbf{4 workers}} \\ \cline{2-7}
   &\multirow{2}{*}{\textbf{TPS}} & \multicolumn{3}{c|}{\textbf{Latency}} & \multirow{2}{*}{\textbf{TPS}} & \textbf{Latency} \\ 
   & & \multicolumn{3}{c|}{\textbf{msec}} & & \textbf{msec} \\ \cline{2-7}
   & \textbf{Obs} & \textbf{Pred} & \textbf{Pred+C+I} & \textbf{Obs} & \textbf{Obs} & \textbf{Obs} \\ \hline
   1 & 6921 & 0.131 & 0.148 & 0.144 & 27091 & 0.148 \\ \hline
   100 & 5246 & 0.159 & 0.189 & 0.191 & 14485 & 0.277 \\ \hline
\end{tabular}
\vspace{2ex}
\caption{TPC-C new-order performance at scale factor 4.}
\label{tab:new-order-validation}
\vspace{-7ex}
\end{table}

We observed in Section~\ref{sec:load:fixed:db} that \textsf{shared-nothing-async} underperforms compared to \textsf{shared-everything-with-affinity} even with a single client worker, despite having more transaction executors available for transaction processing. To better understand this effect, we turned to our cost model for validation of a mix consisting of 100\% new-order transactions with four warehouses and four transaction executors. Similarly to what was done in Section~\ref{sec:exp:pred:programs:breakdown} and Appendix~\ref{sec:exp:pred:limitations}, we calibrated cost model parameters with a new-order transaction requesting only one item from a local and one from a remote warehouse. Moreover, we recorded the average numbers of synchronous and asynchronous stock-update requests realized with a single worker, as well as the average numbers of items involved. To explore the impact of communication overheads, we evaluated two extremes with respectively 1\% and 100\% probability of cross-reactor accesses in stock updates (see also Appendix~\ref{sec:cross:reactor}). 

Table~\ref{tab:new-order-validation} summarizes the results obtained with one and four workers. We obtain excellent fit between the cost model prediction after including commit and input generation cost (Pred+C+I) and the observed latency for one worker under both 1\% and 100\% cross-reactor accesses. 
We observed that the cost of communication between reactors, especially $C_r$ as remarked in Section~\ref{sec:exp:pred:programs}, is high compared with the processing cost of stock updates. However,  
the relatively small growth in the latency for 100\% cross-reactor access with one worker compared with 1\% cross-reactor accesses shows the benefits of overlapping multiple asynchronous sub-transactions across reactors. With four workers, queueing effects manifest with 100\% cross-reactor accesses, which are as discussed in Appendix~\ref{sec:exp:pred:limitations} not part of the cost model, though prediction is still accurate for 1\% cross-reactor accesses.    

The hardware utilization with one worker on the four transaction executor cores under 1\% cross-reactor accesses is 78\%, 2.3\%, 2.3\%, and 2.3\%, respectively, while the values under 100\% cross-reactor accesses are 65\%, 24\%, 24\%, and 24\%. The use of four workers increases utilization, which becomes uniform at 82\% and 86\% under 1\% and 100\% cross-reactor accesses, respectively.  

\begin{figure*}[tp]
  \begin{minipage}{0.32\linewidth}
    \centerline{\includegraphics[width=0.8\linewidth]{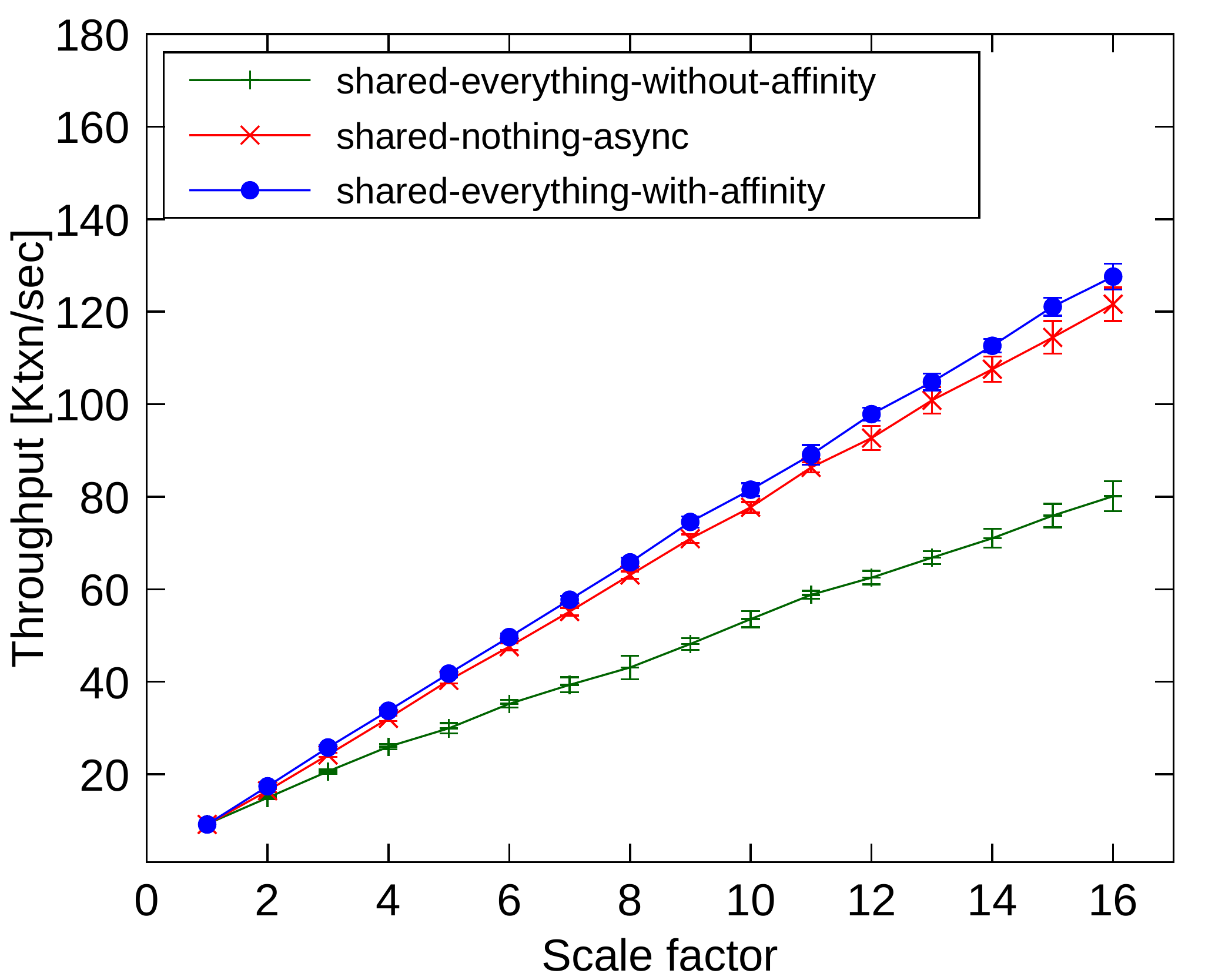}}
    \caption{TPC-C throughput with varying deployments.} \label{fig:throughput:scalability}
  \end{minipage} \hfill
  \begin{minipage}{0.32\linewidth}
    \centerline{\includegraphics[width=0.8\linewidth]{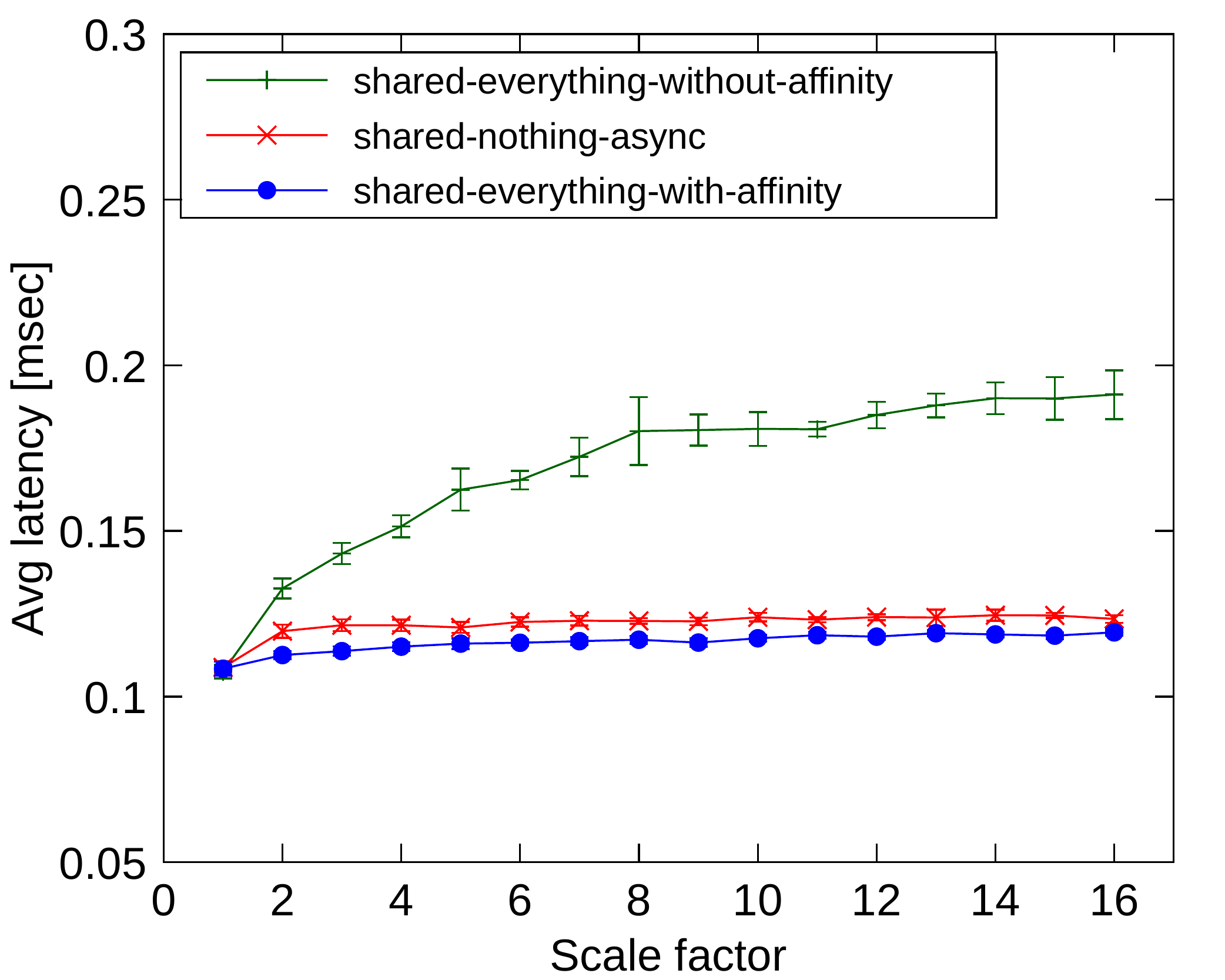}}
    \caption{TPC-C latency with varying deployments.}\label{fig:latency:scalability}
  \end{minipage} \hfill
  \begin{minipage}{0.32\linewidth}
    \centerline{\includegraphics[width=0.8\linewidth]{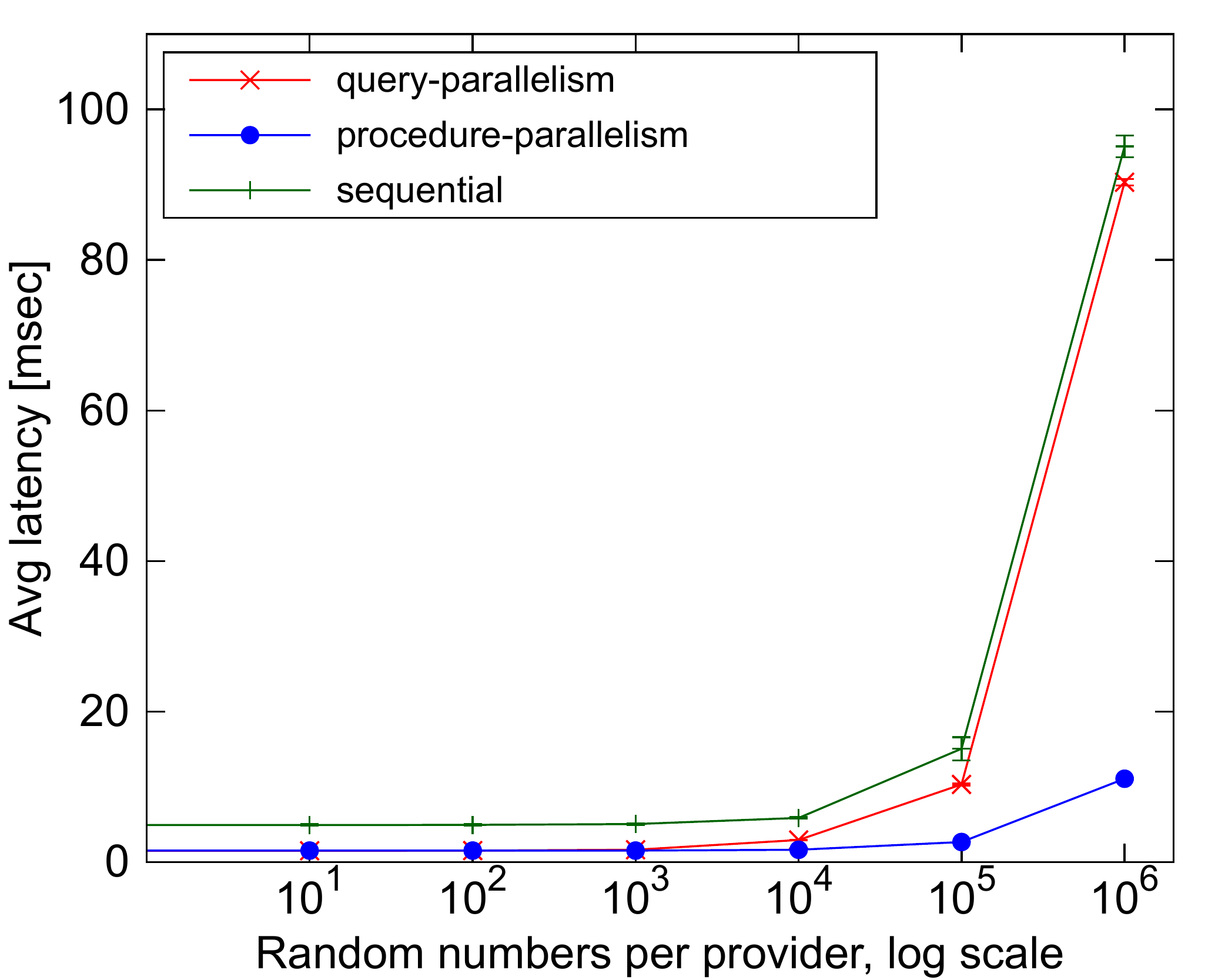}}
    \caption{Latency of query-~vs.~procedure-level~parallelism.}\label{fig:proc:parallelism}
  \end{minipage} \hfill
\end{figure*}

\section{Effect of Cross-Reactor Transactions}
\label{sec:cross:reactor}
In this appendix, we evaluate the impact of cross-reactor transactions on the performance of the different database architectures, complementing the results of Section~\ref{sec:exp:virt}. 
For clarity, we focus exclusively on new-order transactions without any additional computations. To vary the percentage of cross-reactor new-order transactions, we vary the probability that a single item in the transaction is drawn from a remote warehouse (the remote warehouses are again chosen with an equal probability). Each new-order consists of between 5-15 items. 

Figures~\ref{fig:throughput:dist:degree:deployments} and \ref{fig:latency:dist:degree:deployments} show the average throughput and latency, respectively, of running the TPC-C benchmark with 100\% new-order transactions at a scale factor of eight, i.e., eight warehouses receive transactions from eight workers at peak load. Since \reactdb\ uses Silo's OCC protocol and owing to the low contention in TPC-C even upon increasing the number of remote items, we would expect the latency of \textsf{shared-everything-with-affinity} to be agnostic to changes in the proportion of cross-reactor transaction as per the results in~\cite{Tu:2013:STM:2517349.2522713}. However, we see a small gradual 
increase in the latency for all the deployments. We believe these effects are a consequence of cache coherence and cross-core communication overheads. 
We observe further that both \textsf{shared-nothing-sync} and \textsf{shared-nothing-async} configurations exhibit the same latency at 0\% cross-reactor transactions as \textsf{shared-everything-with-affinity}. However, there is a sharp drop in the performance of \textsf{shared-nothing} deployments from 0\% to 10\% cross-reactor transactions. This effect is in line with our previous observation that sub-transaction invocations require expensive migration of control in contrast to both \textsf{shared-everything-without-affinity} and \textsf{shared-everything-with-affinity}. 
 Note that the abort rate for all deployments remained negligible (0.02\%-0.04\%), highlighting the limited amount of contention on actual items.

We observe that \textsf{shared-nothing-async} exhibits higher resilience to increase in cross-reactor transactions when compared with \textsf{shared-nothing-sync}.
In particular, latency of \textsf{shared-nothing-async} is better by roughly a factor of two at 100\% cross-reactor transactions. 
This is because \textsf{shared-nothing-async} employs new-order transactions with asynchronous sub-transaction invocations on remote warehouse reactors, and tries to overlap remote sub-transaction invocation with execution of logic locally on a warehouse reactor. This demonstrates how application programs can leverage the programming model to engineer application code using reactors with different performance characteristics. At the same time, infrastructure engineers can select the database architecture that best fits the execution conditions for the workload without changes to application code. In the case of peak load and limited intra-transaction parallelism, \textsf{shared-everything-with-affinity} turned out to be the best architecture among the ones considered for this scenario, in line with the results of~\cite{Tu:2013:STM:2517349.2522713}.

\section{Scale-Up and Overhead in \reactdb}
\label{sec:scaleup:overhead}
This appendix further complements Section~\ref{sec:exp:virt} by presenting transactional scale-up in \reactdb\ to the degree allowed by our hardware setup and discussing containerization overhead.  

\subsection{Transactional Scale-Up}
In this section, we evaluate the scalability of \reactdb\ across multiple cores for the three database architecture deployments described in Section~\ref{sec:deployments}. 
Figures~\ref{fig:throughput:scalability} and~\ref{fig:latency:scalability} 
show the average transaction throughput and latency, respectively, of running the TPC-C transaction mix as we increase the number of warehouses (reactors). Note that we configure the number of transaction executors to be equal to the scale factor for the experiment.

We observe that the \textsf{shared-everything-without-affinity} deployment exhibits the worst throughput scalability among the deployments selected. This effect is a consequence of \textsf{shared-everything-without-affinity}'s poor ability to exploit memory access affinities within each transaction executor, given round-robin routing of transactions. On the other hand, \textsf{shared-everything-with-affinity} and \textsf{shared-nothing-async} both take advantage of access affinities and behave similarly. We see that \textsf{shared-everything-with-affinity} is slightly superior to \textsf{shared-nothing-async}. The difference lies in the relative costs in these deployments of sub-transaction invocations vs.~direct memory accesses of data for remote warehouses. For a scale factor of one, there are no cross-reactor transactions, and the performance of the two deployments is identical. From a scale factor of two onwards, the probabilities of cross-reactor transactions range between 0\% to 10\%, as discussed in Section~\ref{sec:load:fixed:db}.
In \textsf{shared-nothing-async}, a sub-transaction call is routed to its corresponding transaction executor, incurring context switching and communication overheads. By contrast, since \textsf{shared-everything-with-affinity} executes the sub-transaction in the same transaction executor, the remote call costs are traded off for the relatively smaller costs of cache pressure. We also ran the experiment with all the transaction classes in the TPC-C mix invoking sub-transactions synchronously in the shared-nothing deployment (\textsf{shared-nothing-sync} configuration described in Section~\ref{sec:deployments}). However, the throughput and latency of this configuration was close (within the variance bars) to the \textsf{shared-nothing-async} configuration because of the low percentage of remote warehouse calls in the default TPC-C mix. We hence omit the curve from Figures~\ref{fig:throughput:scalability} and~\ref{fig:latency:scalability} for brevity.

In short, the results indicate that \reactdb\ can be flexibly configured with different database architectures to achieve adequate transactional scalability in our available hardware setup for the standard TPC-C workload. Further,   
high affinity of data accesses to physical processing elements (cores) is crucial to performance.

\subsection{Effect of Affinity}
To further quantify the effect of affinity of reactors to transaction executors on performance, we ran an experiment in which we vary the number of transaction executors deployed in \textsf{shared-everything-without-affinity}, but keep the scale factor of TPC-C at one with a single client worker. In such a setup, for $k$ transaction executors deployed, the load balancing router ensures the $n$-th request is sent to transaction executor $n \mod k$. Thus, the different transactions from the workers are being spread around the transaction executors, which destroys the locality in the transaction executors and accentuates the cache coherence and cross-core communication costs. We found that with two transaction executors throughput drops to 86\% compared to one transaction executor and progressively degrades to 40\% for 16 transaction executors. For comparison, the corresponding per-core throughput for \textsf{shared-everything-with-affinity} at scale factor 16 in Figure~\ref{fig:throughput:scalability} is 87\% of the per-core throughput at scale factor one. This result highlights the importance of maintaining affinity of transaction execution for high performance, especially in a machine with accentuated cache coherence and cross-core communication costs. 

\subsection{Containerization Overheads} 
To account for the overhead of containerization, we also ran \reactdb\ while submitting empty transactions with concurrency control disabled. We observe roughly constant overhead per transaction invocation across scale factors of around 22~$\mu$sec. We measured that thread switching overhead between the worker and transaction executor across different cores is a largely dominant factor and is dependent on the machine used. The overhead accounted for 18\% of average TPC-C transaction latency. When compared with executing the TPC-C application code directly within the database kernel without worker threads generating transaction invocations that are separate from database processing threads, as in Silo,
the overhead is significant, but if a database engine with kernel thread separation is assumed, as is the normal case, the overhead is negligible. 

\section{Procedure-Level Parallelism with Reactors}
\label{sec:procedure:parallelism}

In this section, we revisit the example introduced in Figure~\ref{fig:example:app} and evaluate the potential performance gains that can be obtained by leveraging procedure-level parallelism in the reactor model. In Figure~\ref{fig:example:app}(a), we see the formulation of \texttt{auth\_pay} in the classic transactional model. If the \texttt{orders} relation were partitioned, the join query between \texttt{provider} and \texttt{orders} would be eligible for parallelization by a query optimizer in a traditional relational database. Note that the optimizer would consider query-level parallelization, and not a holistic procedure-level parallelization achievable by manual decomposition of the join in the reactor programming model as shown in Figure~\ref{fig:example:app}(b). In the latter, the potentially expensive computation of \texttt{sim\_risk} would also be parallelized across reactors.

For this experiment, we configured \reactdb\ with up to 16 transaction executors. To focus on intra-transaction parallelism, a single worker generates \texttt{auth\_pay} transaction invocations targeted at one \texttt{Exchange} and 15 \texttt{Provider} reactors, used to express three program execution strategies in \reactdb. Both the \textsf{sequential} and \textsf{query-parallelism} strategies encode the program of Figure~\ref{fig:example:app}(a) for the classic transactional model. In \textsf{sequential}, we deploy a single container and transaction executor for all reactors in \reactdb, and thus the whole procedure body executes in a single hardware thread. In \textsf{query-parallelism}, by contrast, we horizontally decompose \texttt{orders} across the \texttt{Provider} reactors by deploying one container and transaction executor for each reactor. Similar to a partitioned database, this results in a parallel foreign key join between \texttt{providers} and \texttt{orders} fragments, but sequential execution of the remainder of the procedure.  Finally, the \textsf{procedure-parallelism} strategy encodes the program of Figure~\ref{fig:example:app}(b). Using the same deployment as for \textsf{query-parallelism}, this achieves holistic procedure parallelization in the reactor model. The inputs for \texttt{auth\_pay} were chosen using uniform distributions, and the \texttt{orders} relation was loaded with 30,000 records per provider, mirroring the cardinality of the \texttt{orders} relation in the TPC-C benchmark. To simulate another transaction settling orders and keep the number of records scanned in the \texttt{orders} relation fixed, a pre-configured window of records is reverse range scanned ordered by time per provider. This window parameter has a direct effect on the benefit of \textsf{query-parallelism} compared with \textsf{sequential}. For clarity, we tuned this value to 800 records per provider to ensure that \textsf{query-parallelism} would outperform \textsf{sequential} by 4x when \texttt{sim\_risk} is not invoked. We simulated the computational load of \texttt{sim\_risk} by random number generation similar to Section~\ref{sec:load}. We also loaded our data values such that \texttt{sim\_risk} is always invoked and transactions are not aborted due to application logic failures. 
  
Figure~\ref{fig:proc:parallelism} shows average latencies as we vary the computational load in \texttt{sim\_risk} by generating progressively more random numbers. We observe that \textsf{procedure-parallelism} is more resilient to changes in computational load and better utilizes the available parallel hardware in this benchmark. At the extreme of $10^6$ random numbers generated per provider, the latency of \textsf{procedure-parallelism} is a factor of 8.14x and 8.57x lower than that of \textsf{query-parallelism} and \textsf{sequential}, respectively. The transaction executor core hosting the \texttt{Exchange} reactor becomes overloaded at 100\% utilization under \textsf{query-parallelism}, while with \textsf{procedure-parallelism} utilization remains uniform at 84\% across executor cores hosting \texttt{Provider} reactors and 14\% in the \texttt{Exchange} executor core. In \textsf{sequential}, the transaction executor mapped to all reactors becomes overloaded at 100\% even when the computational load is zero.

\section{Smallbank Reactor Implementation Details}
\label{sec:smallbank:reactor:implementation}
\begin{figure}[!t]
        \begin{minipage}{1.0\columnwidth}
          \centerline{\includegraphics[width=0.35\linewidth]{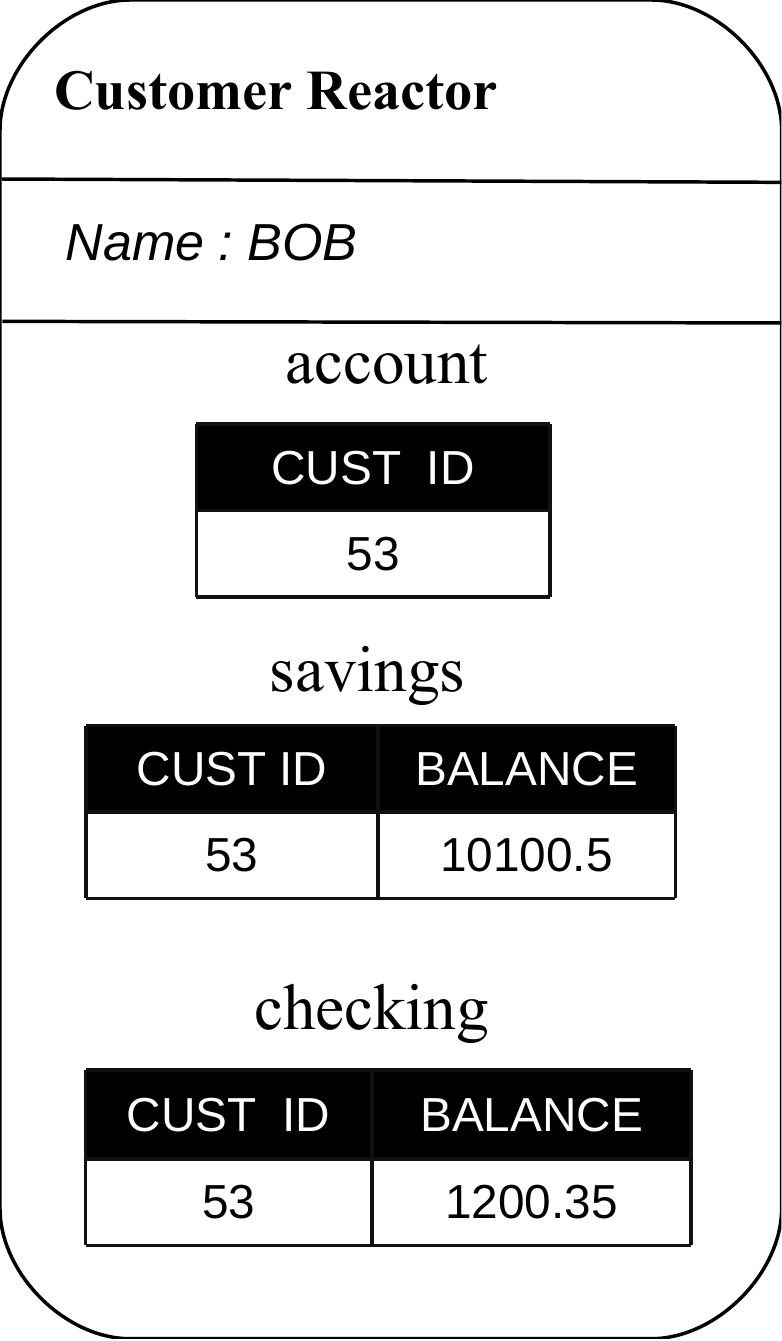}}
          \caption{Example Customer Reactor in Smallbank.}
          \label{fig:example:smallbank}
        \end{minipage} \hfill
\end{figure}

\begin{figure}[!t]
        \begin{minipage}{1.0\columnwidth}
          \centerline{\includegraphics[width=0.91\linewidth]{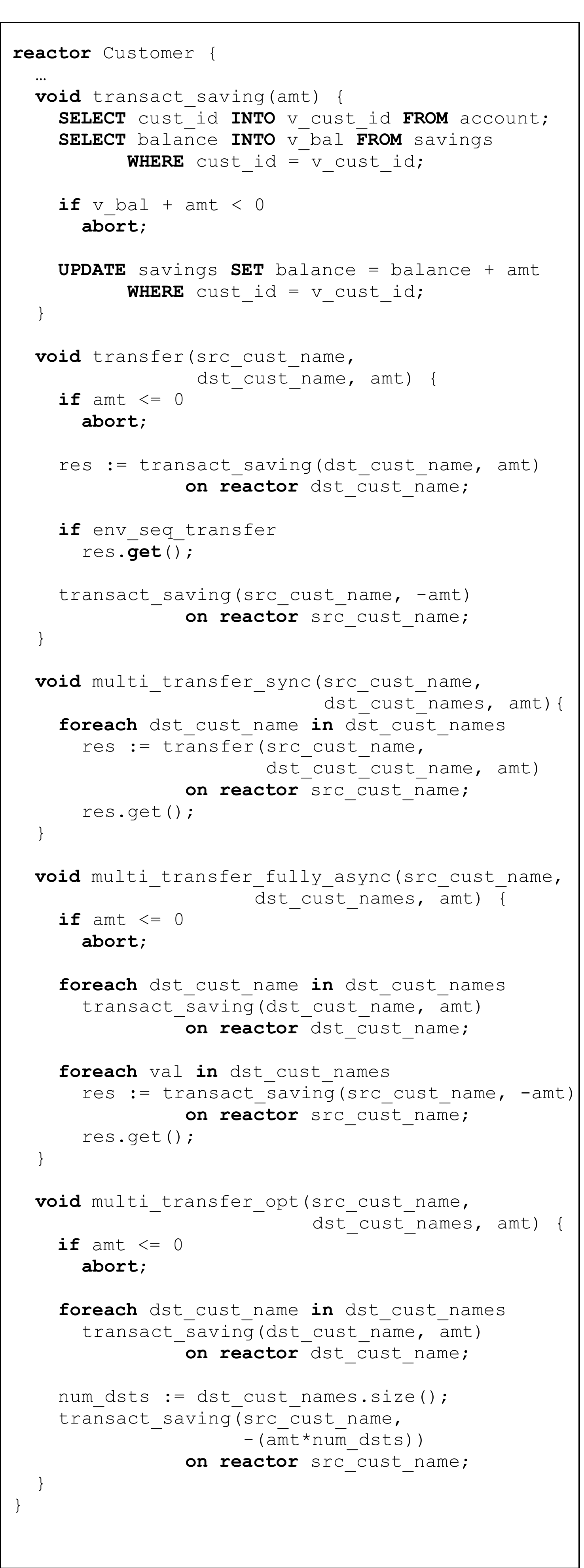}}
          \caption{Implementation of Smallbank multi-transfer transactions.}
          \label{fig:example:smallbank:code}   
        \end{minipage} \hfill
\end{figure}

In this appendix, we provide implementation details of the application programs in the Smallbank benchmark, which were used in the experiments in Section~\ref{sec:exp:pred} and Appendix~\ref{sec:exp:pred:cost}. In this benchmark, each customer was modeled as a reactor. Figure~\ref{fig:example:smallbank} outlines the encapsulated relations on each \texttt{Customer} reactor, namely (1) \texttt{account}, which maps the customer name to a customer ID, (2) \texttt{savings} and (3) \texttt{checking}, which represent the savings and checking accounts of the customer, respectively. For strict compliance with the benchmark specifications, we have maintained the customer ID field in the savings and checking relations despite it being a relation holding a single tuple for a customer reactor. We have also performed the lookup on the \texttt{account} relation for customer ID followed by its use on the \texttt{saving} and \texttt{checking} relation to maintain the logic and query footprint of the benchmark specification. Figure~\ref{fig:example:smallbank:code} shows the various methods exposed by the customer reactor to represent the various formulations of the multi-transfer application logic.

\begin{sloppypar}
The \texttt{transfer} transaction uses an enviroment variable during compile time (\textbf{\texttt{env\_seq\_transfer}}) that can be enabled or disabled to execute the \texttt{multi\_transfer\_sync} method in \textsf{fully-sync} or \textsf{partially-sync} mode.\footnote{This also helps in minimizing code duplication.} In the benchmark, \texttt{multi\_transfer\_sync}, \texttt{multi\_transfer\_fully\_async}, and \texttt{multi\_transfer\_opt} were invoked on the source customer reactor from which the amount must be transferred to the destination customer reactors. 
The explicit synchronization in \texttt{multi\_transfer\_sync} is done for safety, though not required when the \texttt{src\_cust\_name} customer reactor executes the method. This is because in this case the nested \texttt{transact\_saving} sub-transaction on the \texttt{src\_cust\_name} reactor is executed synchronously producing the same effect. 
However, explicitly specifying the synchronization there improves code clarity. 
For the same reason, explicit synchronization on the \texttt{transact\_saving} sub-transaction is done in the \texttt{multi\_transfer\_fully\_async} method as well.
\end{sloppypar}

\end{document}